\documentclass[aps,pre,onecolumn,superscriptaddress,groupedaddress,showkeys]{revtex4} 

\usepackage{amsmath}
\usepackage{amssymb}
\usepackage{amsthm}
\usepackage{mathrsfs}
\usepackage{mathtools}
\usepackage{bm}
\usepackage{xcolor}
\usepackage{array}
\usepackage{algpseudocode}
\usepackage{algorithm}

\newcommand{\R}{\mathbb{R}}                                     
\newcommand{\X}{\mathbb{X}}                                     
\newcommand{\Y}{\mathbb{Y}}                                     
\newcommand{\M}{\mathbb{M}}

\newcommand{\ts}{\hspace*{0.1em}} 

\begin{document}

\title{Collective variables between large-scale states in turbulent convection}

\author{Priyanka Maity}
\affiliation{Department of Mechanical Engineering, Technische Universit\"at Ilmenau, Postfach 100565, D-98684 Ilmenau, Germany}
\author{Andreas Bittracher}
\affiliation{Department of Mathematics, Freie Universität Berlin, Arnimallee 6, D-14195 Berlin, Germany}
\author{P\'eter Koltai}
\affiliation{Department of Mathematics, University of Bayreuth, Universitätsstraße 30, D-95440 Bayreuth, Germany}
\affiliation{Department of Mathematics, Freie Universität Berlin, Arnimallee 6, D-14195 Berlin, Germany}
\author{J\"org Schumacher}
\affiliation{Department of Mechanical Engineering, Technische Universit\"at Ilmenau, Postfach 100565, D-98684 Ilmenau, Germany}
\affiliation{Tandon School of Engineering, New York University, New York, NY 11201, USA}


\date{\today}

\begin{abstract}
The dynamics in a confined turbulent convection flow is dominated by multiple long-lived macroscopic circulation states which are visited subsequently by the system in a Markov-type hopping process. In the present work, we analyze the short transition paths between these subsequent macroscopic system states by a data-driven learning algorithm that extracts the low-dimensional transition manifold and the related new coordinates, which we term collective variables, in the state space of the complex turbulent flow. We therefore transfer and extend concepts for conformation transitions in stochastic microscopic systems, such as in the dynamics of macromolecules, to a deterministic macroscopic flow. Our analysis is based on long-term direct numerical simulation trajectories of turbulent convection in a closed cubic cell at a Prandtl number $Pr = 0.7$ and Rayleigh numbers $Ra = 10^6$ and $10^7$ for a time lag of $10^5$ convective free-fall time units. The simulations resolve vortices and plumes of all physically relevant scales resulting in a state space spanned by more than 3.5 million degrees of freedom. The transition dynamics between the large-scale circulation states can be captured by the transition manifold analysis with only two collective variables which implies a reduction of the data dimension by a factor of more than a million. Our method demonstrates that cessations and subsequent reversals of the large-scale flow are unlikely in the present setup and thus paves the way to the development of efficient reduced-order models of the macroscopic complex nonlinear dynamical system. 
\end{abstract}


\maketitle

\section{Introduction}
Complex nonlinear systems typically incorporate orders of magnitude of relevant dynamical scales. Examples start at the microscopic {\em stochastic} level where protein macromolecules, that remain in a certain conformation for milliseconds, switch within nanoseconds into a different configuration that leads to a significant change in their chemical functionality \cite{jackson1998,rose2006}. All the way up to the macroscopic {\em deterministic} level, turbulent flows in confined geometries or extended layers can exhibit differently ordered large-scale spatial patterns which are visited for longer transients in a long-term evolution ~\cite{spiegel_1971,markson_1975,marshall_schott_1999,schumacher_2020}. The (rapid) crossover from one configuration to another is triggered by fluctuations of secondary flow structures, smaller eddies, shear layers or plumes that can affect the turbulent transport of heat or momentum \cite{xi_2008,vanderpoel_2008,weiss_2011,huisman_2014,bao_2015,zwirner_2020,schindler_2022}. The state or phase space of macroscopic flows is infinite-- or at least extremely high--dimensional and requires drastic dimensionality reductions to model the observed large-scale dynamics effectively~\cite{ruelle_1995}. 

The present study transfers an unsupervised data reduction strategy, the transition manifold framework, from stochastic molecular dynamics \cite{schultze_2021, BitEtAl17, bittracher_2021} to deterministic turbulent flows. More precisely, we will analyze a turbulent thermal convection flow, also denoted as Rayleigh--B\'{e}nard convection (RBC) flow, which is driven by buoyancy forces and is confined in a closed cubic cell~\cite{chandrashekhar_1961,ahlers_2009,chilla2012}. Despite the simple dynamics and physical-space geometry, this configuration serves as a paradigm for many applications in nature and technology. The present turbulent flow has several similarities with the mentioned microscopic example. It appears in different conformations, here different large-scale circulation (LSC) states \cite{niemela_2001, sreenivasan_2002, parodi_2004, puthenveettil_2005, brown_2006, xi_2007} which occupy different regions of the phase space for many convective time units before rapidly switching from one to another in phase space. On the one hand, one could think of these LSC states to correspond in phase space to states marked by strong similarity in their velocity and temperature fields, hence they can be thought to build concentrated clusters. On the other hand, the transitory dynamics between the LSC states may be arbitrarily complicated, providing a---if not \emph{the}---major obstacle to reduced modelling of this system. It is exactly the transitory dynamics, more precisely the progress between the different LSC states, that we will target here. In particular, we will provide data-based evidence that it is low-dimensional.

The LSC in a closed cubic cell appears in the form of four diagonal discrete circulation roll states which fill the whole cell aligned along the diagonal~\cite{foroozani_2014, bai_2016, foroozani_2017, giannakis_2018, vasilev_2019}. It was shown recently that this hopping from one long-lived LSC (LL-LSC) state to another, which proceeds via short-lived LSC (SL-LSC) states, can be approximately described as a Markov process based on an analysis of LL-LSC lifetimes and transition probabilities~\cite{Maity_2021}.

The analysis of the fast transition events of the long-term trajectory will give a minimal set of new coordinates spanning a low-dimensional \emph{surrogate state space} that is sufficient to represent the statistical dynamical behavior of the system. The description by these new coordinates is then connected to the dynamical processes in the turbulent RBC flow in the physical space, namely the intermittent interplay of local corner vortices next to the LSC which grow transiently and kick the diagonal circulation roll into the next macrostate. Our approach thus opens the door to an efficient reduced dynamical model of the LSC dynamics in this particular application, namely as a Markov process on the surrogate state space.

These new coordinates are often termed in the original chemical context as reaction coordinates~\cite{BitEtAl17}, we will use the notion of {\em collective variables} for the present application. The turbulence data stem from a long-term simulation trajectory of $10^5$ convective time units $t_f$ at two different Rayleigh numbers $Ra=10^6,\, 10^7$ and a Prandtl number $Pr=0.7$, see the next section for the Rayleigh--B\'{e}nard flow model of thermal convection and exact definitions of $Ra$ and~$Pr$. Given the spectral resolution in the simulations for the 4 turbulence fields invovled, we count more than 3.5 million degrees of freedom that describe the turbulent convection flows. 

Recent years have witnessed a large bloom in designing methods that use (deep) neuronal networks to find surrogate dynamical models and in particular low-dimensional variables in terms of which these models are expressed~\cite{brandt2018machine,wehmeyer2018time,lusch2018deep,mardt2018vampnets,champion2019data,otto2019linearly}. Their success relies strongly on the ability of such networks to represent coordinates from a large general class. In these approaches, however, the dynamical conditions necessary for them to perform well remain implicit. The methodology presented here relies on very explicit dynamical assumptions, that are validated in course of the data-driven computation.

\section{Turbulent Rayleigh--B\'{e}nard flow}
\label{DNS}

\subsection{Model equations and simulation method} 
We simulate the dimensionless three-dimensional incompressible Boussinesq equations of motion representing the Rayleigh--B\'enard convection dynamics. They are given by
\begin{align}
\partial_t {\bf u} + ({\bf u} \cdot \nabla) {\bf u} &= -\nabla p + \sqrt{\frac {Pr}{Ra}} \nabla^2 {\bf u} +  T \hat{z}\,,\label{NS1} \\
\partial_t T + ({\bf u} \cdot \nabla) T &= {\frac{1}{\sqrt{Ra Pr}}} \nabla^2 T\,, \label{temp1}\\
\nabla \cdot {\bf u} &= 0\,.\label{cont1}
\end{align}
Here, ${\bf u}({\bf x},t)$ is the velocity field of the fluid, $T({\bf x},t)$ the temperature field, and $p({\bf x},t)$ the pressure field. The dimensionless control parameters of the flow are the Rayleigh number $Ra$, which gives a measure of the strength of driving by buoyancy forces and Prandtl number $Pr$, which is the ratio of momentum to thermal diffusivity. Both numbers are given by 
\begin{equation}
Ra = \frac{\alpha g\, \delta T d^3}{\nu \kappa} \quad\mbox{and}\quad
Pr=\frac{\nu}{\kappa}\,,
\end{equation}
where $\alpha$ represents the isobaric thermal expansion coefficient, $g$ the acceleration due to gravity, $\delta T=T_{\rm bottom}-T_{\rm top}$ the temperature difference maintained along the fluid layer of thickness $d$, and $\nu$ and $\kappa$ the kinematic viscosity and thermal diffusivity of the fluid, respectively. The equations have been made dimensionless by re-scaling lengths by the height of the cell $d$ (which is equal here to the two horizontal side lengths), velocities by the free-fall velocity $U_f = \sqrt{\alpha g \delta T d}$, and temperatures by the outer difference $\delta T$. The time units for non-dimensionalization results then to the free-fall timescale $t_f = \sqrt{d/(\alpha g \delta T)}$ which is the large-scale convective time unit in the present work. 

We assume no-slip velocity boundary conditions at all six faces of the cubic cell $V=d^3$ with ${\bf u} = 0$. The system is uniformly heated from below at $z=0$ with $T=T_{\rm bottom}$ and cooled from above at $z=1$ with $T=T_{\rm top}$. We also assume thermally insulated side faces, ${\bf n}\cdot \nabla T=0$, such that all supplied heat at the bottom has to pass through the fluid to the top.

The simulations were performed using the open source code Nek5000 (version 17) which is based on a spectral element method \cite{fischer_1997,scheel_2013}. The simulations were performed assuming 16 spectral elements along each space direction and a Lagrangian interpolation polynomial of order 5 (for $Ra=10^6$) and 7 (for $Ra=10^7$) along each space direction and on each spectral element, which results in 884,736 and 2,097,152 collocation points, respectively. The vertical profiles of the mean kinetic energy dissipation rate were analyzed to verify that this spectral resolution is sufficiently well resolved for analyzing the large scale circulations. The simulations were carried out for two values of Rayleigh numbers $Ra = 10^6$ and $10^7$ in a fluid with Prandtl number $Pr=0.7$, corresponding to an effective Reynolds number of $Re=\sqrt{Ra/Pr}\,u_{\rm rms}$ of 467 and 1350 respectively, with $u_{\rm rms}$ denoting the root-mean-square velocity, see table \ref{tab2}. Starting from a random initial condition, we waited for a lag of 5000 free-fall times for the system to settle to a steady state. The trajectory was further allowed to evolve for another $10^5$ free-fall times, and the turbulence fields were output at every free-fall time $t_f$, such that we gather $N_s=10^5$ full flow snapshots for each case considered here. 
\begin{table}
\renewcommand{\arraystretch}{1.2}
\begin{tabular}{lcccc} 
\hline\hline
$(Pr, Ra)$     & $\quad(0.7, 10^6)\quad$ & $(0.7, 10^7)$ \\
     \hline
$Re$ & 467 & 1350  \\
$u_{\rm rms}=\langle u_i^2\rangle_{V,t}^{1/2}$ & 0.39 & 0.36\\
$\omega_{\rm rms}=\langle \omega_i^2\rangle_{V,t}^{1/2}$ & 2.86 & 3.83\\
$\theta_{\rm rms}=\langle \theta^2\rangle_{V,t}^{1/2}$ & 0.29 & 0.29\\
$t_{\rm pers}$ (SL-LSC) &4.2$t_f$ & 8.6$t_f$  \\
$t_{\rm pers}$ (LL-LSC) & 12.1$t_f$ & 26$t_f$ \\
\hline\hline
\end{tabular}
\caption{\label{tab1} Summary of basic parameters and turbulence quantities. Comparison of root-mean-square values of velocity ($u_{\rm rms}$), vorticity ($\omega_{\rm rms}$), and temperature ($\theta_{\rm rms}$) obtained for various combinations of Rayleigh number $Ra$ and Prandtl number $Pr$. Here, $\langle\cdot\rangle_{V,t}$ is a combined average over the cubic volume and time. The mean persistence times ($t_{\rm pers}$) of the long-lived LSCs (LL-LSC) and short-lived LSCs (SL-LSC) are enlisted in the last two rows in terms of convective free-fall time units $t_f$. They are obtained from the fit of exponential laws to the persistence time distributions.}
\label{tab2}
\end{table}
\begin{figure}
\centering\includegraphics[width=0.45\textwidth]{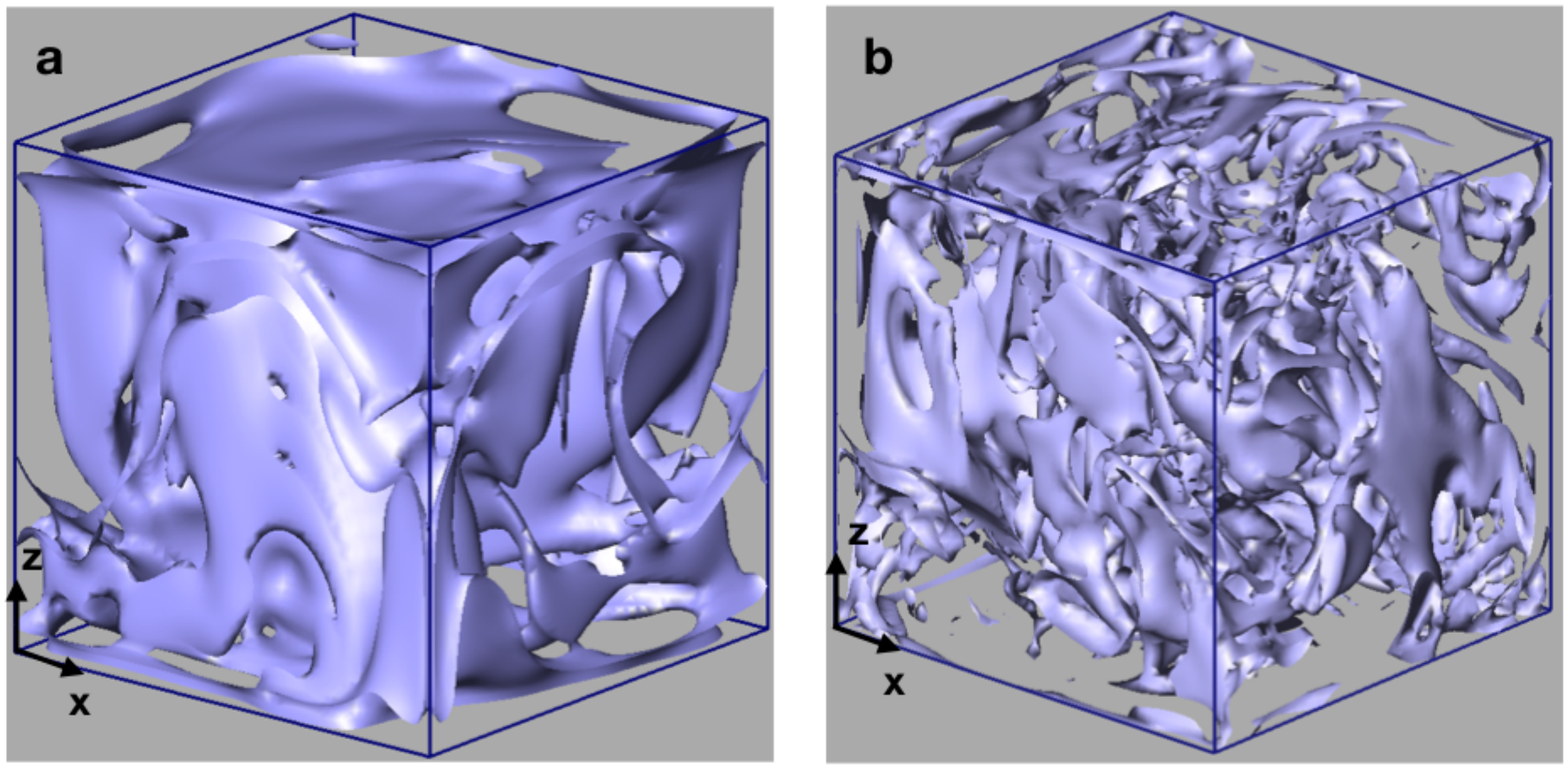}
\caption{Iso-surfaces of vorticity magnitude plotted for the case of fluid Pr = 0.7 at Rayleigh numbers of (a) $Ra=10^6$ and (b) $Ra=10^7$. Displayed are two time instants of the convection flows. The iso-surfaces are plotted at a value of $|{\mathbf \omega}| = 1.5$, which corresponds approximately to the most probable vorticity magnitude in both flows.} 
\label{vor_dist}
\end{figure}
\subsection{Velocity and vorticity statistics} 
\label{subsec:stat_analysis}
Table~\ref{tab2} lists the velocity, vorticity, and temperature fluctuations in the system for both runs. The fluctuations are determined as a root mean square with respect to the flow volume $V$ and time $t$. Our data analysis revealed that a lower switching frequency between the long-lived LSC states is observed, even though  a higher level of vorticity fluctuations in the convection flow exists when comparing the two long-term runs at $Ra=10^6$ and $10^7$ and $Pr=0.7$. As also seen, the root mean squared values of the velocity, $u_{\rm rms}$ and temperature fields, $\theta_{\rm rms}$, remain nearly unchanged. 

The higher magnitude of vorticity fluctuations $\omega_{\rm rms}$ at the higher Rayleigh number implies that the vortical structures are fragmented and thinner for the higher $Ra$, whereas we observed more coherent vortical structures for the lower one. This is confirmed by the plots of vorticity isosurfaces in Fig.~\ref{vor_dist}. The absence of spatially coarse coherent vortical structures in case of $Ra=10^7$ might explain the longer lifetime of the LSC structures in one specific macrostate. Our observation is in accordance with previous studies in cylindrical cells at the same aspect ratio~\cite{scheel_schumacher_2016,scheel_schumacher_2017}. Following ref.~\cite{scheel_schumacher_2017}, the dependence of the mean kinetic energy dissipation rate in turbulent convection at $Pr=0.7$ on the Rayleigh number follows the scaling law
\begin{equation}
    \langle\epsilon\rangle_{V,t}\sim Ra^{-0.2}
\end{equation}  
with the kinetic energy dissipation rate field
\begin{equation}
\epsilon({\bf x},t)=\sqrt{\frac{Pr}{4Ra}}\left[{\bm \nabla u}+({\bm \nabla u})^T\right]^2\,. 
\label{ediss}
\end{equation}
The mean kinetic energy dissipation rate is in turn to a very good approximation directly proportional to the mean square of the vorticity, as already found in \cite{scheel_schumacher_2016,pandey_physD_2022}. This implies 
\begin{equation}
    \langle\epsilon\rangle_{V,t}\approx \sqrt{\frac{Pr}{Ra}} \omega_{\rm rms}^2  \sim Ra^{-0.2}\,,
\end{equation}
and thus a growth with $Ra$ following $\omega_{\rm rms} \sim Ra^{0.15}$. Note that this is an exact equality in homogeneous isotropic turbulence; the proportionality constant is the kinematic viscosity.

\section{Large-scale circulation}
\label{sec:LSC}

\subsection{Detection of large-scale circulation states}
\label{subsec:detection_LSC}

In the following, we discuss how the large-scale flow states are obtained from the numerical simulation data. For this analysis, the scalar and vector fields were interpolated spectrally to a uniform Cartesian mesh. The visualization and computation of the time evolution of the angle of orientation of the LSC followed that of \cite{Maity_2021}. Subsequently, we interpolate the uniform grid data of the vertical velocity component to a circle with a fixed radius of $r=0.45$ and an angle $\theta$ varying in steps of $5^{\circ}$. The angle $\theta$ is measured with respect to the $y$-axis in a clockwise manner. We then performed an azimuthal Fourier transform of the vertical velocity component data interpolated on the circle. The identification of orientation angle $\theta$ of the LSCs is  conditioned on two quantities: (i) The ratio of energy carried by the largest Fourier mode, which is given by
\begin{equation}
\delta(t)=\frac{\max_{k_{\theta}} |\hat{u}_z(k_{\theta},t)|^2}{\sum_{k_{\theta}} |\hat{u}_z(k_{\theta},t)|^2}\,,
\end{equation}
and (ii) the phase of the largest Fourier mode. During the time intervals with a pronounced LSC structure, the maximum energy will be possessed by this particular Fourier mode, and we will have a large value of $\delta(t)$. The orientation of the LSC structure corresponds to the phase of the largest Fourier mode. In absence of any LSC structure, the kinetic energy will be distributed amongst all the modes, and we will have a low value of $\delta(t)$. The decoherent states are extremely rare and unstable. For a methodical approach of identifying LSCs, we first calculated the standard deviation $\sigma$ of the PDF of $\delta$. All flow states which fall below $\mu-2\sigma$ (with $\mu$ being the mean of the PDF) and thus have extremely low values of $\delta$ are identified as decoherent states. All remaining states having values greater than $\mu + 2\sigma$, are categorized as one of the four distinct LSC structures with an angle of orientation $\theta$ corresponding to the phase of the largest Fourier mode.  

\begin{figure*}[ht]
    \centering
    \includegraphics[width=0.9\textwidth]{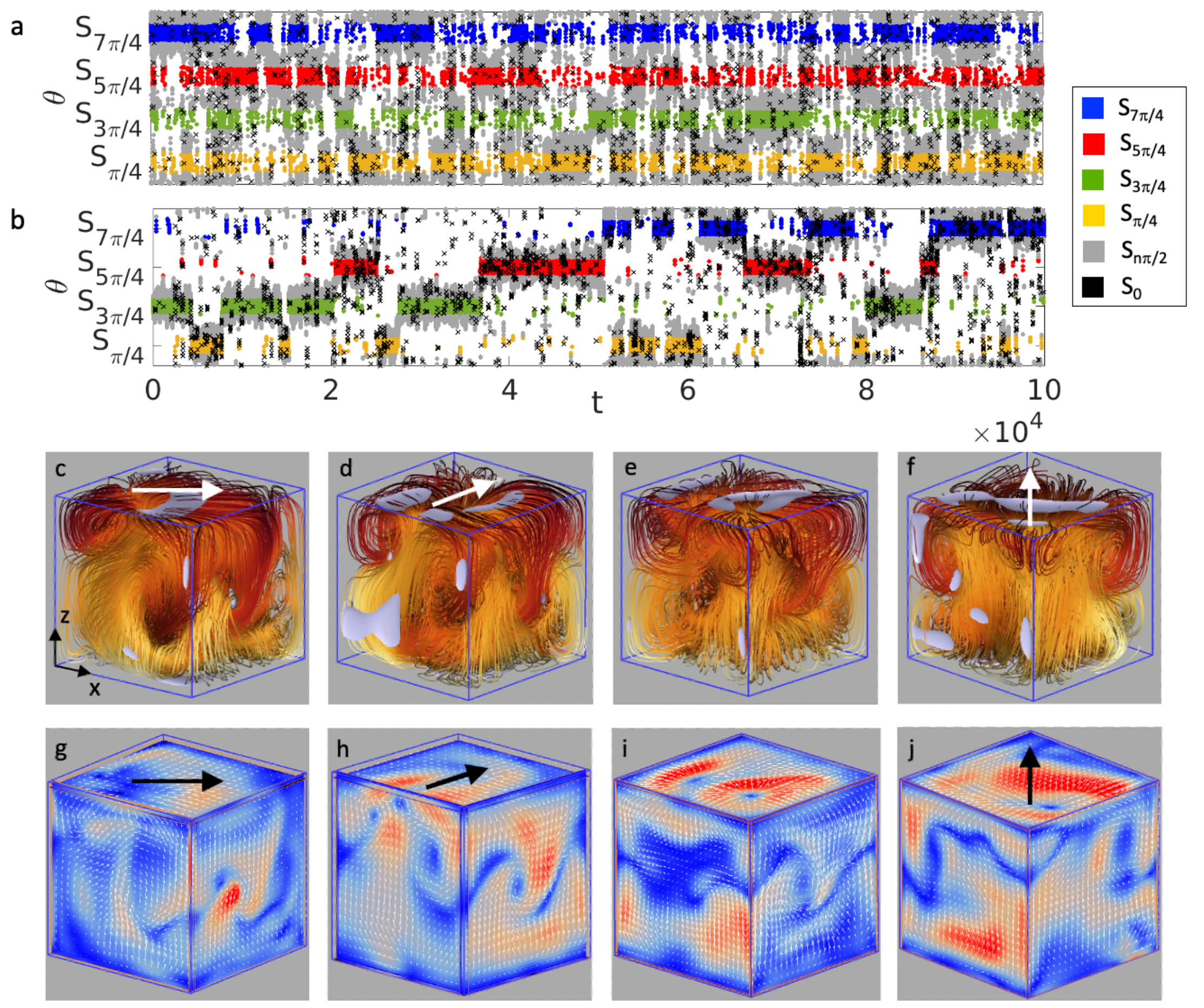}
    \caption{Large-scale flow behavior of the convection flow in the cubical cell. Temporal evolution of orientation angle $\theta$ of the large-scale circulations (LSCs) for Prandtl number $Pr = 0.7$ and two different Rayleigh numbers. (a) Ra = $10^6$. (b) Ra = $10^7$. The legend for both panels indicates the six different LSC states. (c--j) Volumetric visualization of an example of a fast transition between different LSC states. Panels (c)--(f) represent velocity streamline plots of the LSC states in the course of a transition from the long-lived LSC state S$_{5\pi/4}$ in panels (c,g) to the intermediate short-lived LSC state S$_{n\pi/2}$ state in (d,h) and via the decoherent null state S$_0$ in (e,i) to the new long-lived LSC state S$_{3\pi/4}$ state in (f,j). The velocity streamlines are colored according to the temperature at that instant with yellow denoting highest and black representing the lowest temperature, respectively $(0\le T\le 1)$. The direction of the large-scale circulation flow is indicated by white arrows, except for the case of null states. The isosurfaces of the vorticty magnitude with $|{\bm \omega}|\ge 7$ are added. Corresponding velocity vectors are plotted in panels (g)--(j), where blue and red contours represent the minimal, $|{\bm\omega}|=0$, and maximal vorticity magnitudes, $|{\bm \omega}|=8$, respectively. The direction of alignment of LSCs are represented again by black arrows.}
    \label{fig:1}
\end{figure*}

We identify the six different discrete LSC states of the convection flow heuristically on the basis of the energy content contained in Fourier modes. Note that this classification heavily relies on our prior knowledge of the geometry of the physical space and the already observed LSC configurations. We observe that the system mainly prefers the long-lived large scale circulations (LL-LSC) states aligned along the diagonals, termed as S$_{\pi/4}$, S$_{3\pi/4}$, S$_{5\pi/4}$, and~S$_{7\pi/4}$. The transitions between the LL-LSC states occur via the short-lived LSC (SL-LSC) states termed S$_{n\pi/2}$, which are aligned along the side faces, and the decoherent state~S$_0$. We monitored the large-scale state every convective free-fall time based collecting a total of $10^5$ full three-dimensional convection flow snapshots for each of the two simulation runs. The evolution of the long-term trajectories of the LSC orientation $\theta$ for both runs are shown in Fig.~\ref{fig:1}~(a,b). It is noted that the transition frequency in the case of $Ra=10^7$ is lower than for $Ra=10^6$. This has also been observed in quasi-two-dimensional flows~\cite{sugiyama_2010}. The lower level of angular fluctuations results in the impression that there are significantly less data points in the bottom panel; the number of data points is $N_s=10^5$ in both cases.

One such transition between two long-lived LSC states is detailed in Fig.~\ref{fig:1}~(c--j) as a typical example, the switch from the long-lived S$_{5\pi/4}$ to the long-lived S$_{3\pi/4}$ via the short-lived S$_{n\pi/2}$ and the decoherent null state S$_0$ proceeds within a relatively short time period of 18 convective free-fall times. This is about the time it takes for a fluid parcel to circulate within an LSC roll; thus it is considered as a fast process in comparison to the total time lag. Three-dimensional streamline views from the side superimposed with iso-surfaces of high-vorticity magnitudes (in gray) are shown in the middle row of the figure. We clearly observe that these intense vorticity structures are always present on both sides of the LSC mean flow, see panels (c)--(f). Their imbalance in terms of the magnitude should thus be responsible for the destabilization of the LSC roll. The significance of high-vorticity regions for large-scale flow cessations is known from quasi-two-dimensional cases~\cite{sugiyama_2010}; the composition of corner vortices is however different in~3D. A visualization of vorticity magnitude contours in the bottom row of the figure in panels (g)--(j) reveals that the high-vorticity regions are a part of the curling arms of the rising plumes. The direction of the alignment of the LSC roll is determined by interactions between three-dimensional up- and down-welling plumes in the system, see, e.g., the rising plume in the centre of the left front face and the falling one in the right front face of Fig.~\ref{fig:1}~(d,h). In the section~\ref{subsec:stat_analysis}, we also show that the magnitude of vorticity fluctuations increases with increasing Rayleigh number and that the vortical structures become more filamented, see Fig.~\ref{vor_dist} and table~\ref{tab2}. This example also demonstrates that a complex fully three-dimensional pattern is incorporated into the switching process, e.g.,~S$_0$. A description of this dynamics by a minimal set of coordinates will be essential to build a reduced order model of the large-scale flow behavior. 


\subsection{Transition between large-scale circulation states}

\begin{figure}[htb]
\centering\includegraphics[width=0.46\textwidth]{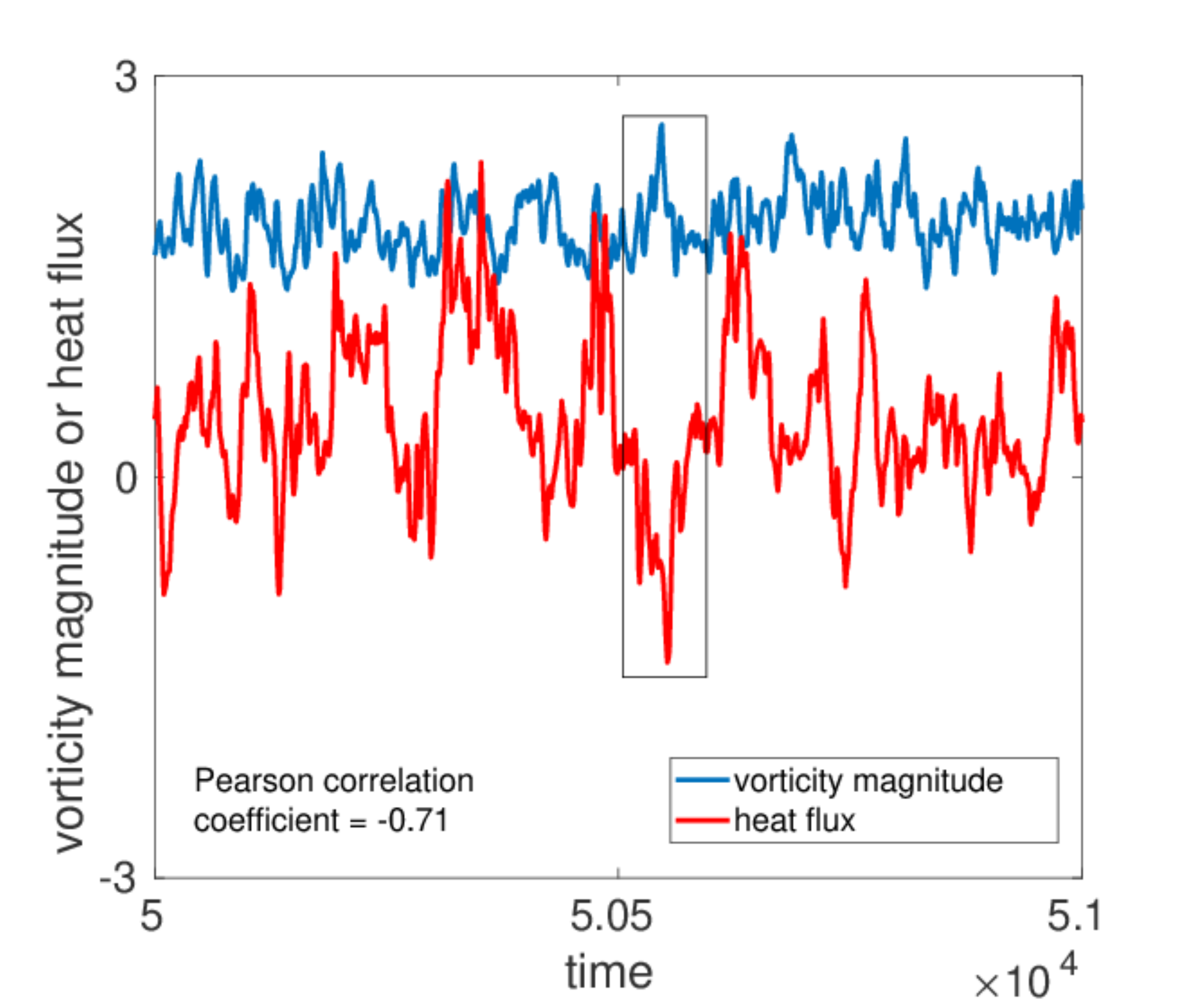}
\caption{Transition from LSC state S$_{5\pi/4}$ to S$_{7\pi/4}$ via a short-lived state for~$Ra = 10^7$ monitored by a comparison between sub-volume-averaged convective heat flux $\overline{\langle u_z T(t)\rangle}_s$ computed along the direction of alignment of the LSC, and the sub-volume-averaged vorticity magnitude $\overline{\langle |{\bm \omega}|(t)\rangle}_s$ computed in the opposite corner sub-volumes where the intense vortical structures reside. The convective heat flux is also scaled up by a multiplicative factor of 50 for better visual comparison with the vorticity magnitude. The destabilization of long-lived LSC occurs when the vorticity peaks and the heat flux minimizes. One such destabilization event occurs in the boxed time window of 100 free-fall times. The Pearson correlation coefficient between both time signals in this window is $-0.71$ which highlights the strong anti-correlation near the transitions.} 
\label{trans_mech}
\end{figure}
Our observation suggests that the long-lived LSC states always destabilize to S$_{n\pi/2}$ (short-lived states) or S$_{0}$ (decoherent state) before converting into another long-lived LSC state. To substantiate our proposition of the  destabilization mechanism of the LSC by the coherent vortices in the corners along the plane perpendicular to the direction of circulation, we compare the heat flux along the flow alignment and the vorticity magnitude triggering the destabilization.  For this purpose, we extracted sub-volumes $V_s=(d/4)^3$, which are centered in the eight corners of the cubic simulation box. Thereafter, we calculate the vorticity centers within each of the eight $V_s$ which are given by
\begin{equation}
    {\bf r}_s(t) =\dfrac {\int_{V_s} |{\bm \omega}({\bf x},t)| {\bf x}\, dV} {\int_{V_s} |{\bm \omega}({\bf x},t)|\, dV}\,.
    \label{vorticity_centers}
\end{equation}
At these eight vorticity centers ${\bf r}_s(t)$ (which are the ``center of mass`` of high-vorticity regions in $V_s$), we center a further smaller cubic sub-volume $V_c$ consisting of $8^3$ grid points to calculate the following time series
\begin{equation}
    \langle|{\bm \omega}|(t)\rangle_s =\frac{1}{V_c} \int_{V_c} |{\bm \omega}({\bf x}+{\bf r}_s(t),t)|\, dV\,.
    \label{vorticity_centers1}
\end{equation}
Similarly, we proceed for the convective heat flux,
\begin{equation}
    \langle u_zT(t)\rangle_s =\frac{1}{V_c} \int_{V_c} u_z({\bf x}+{\bf r}_s(t),t) T({\bf x}+{\bf r}_s(t),t)\, dV\,.
    \label{vorticity_centers2}
\end{equation}
Finally, we take an arithmetic average $\bar{(\cdot)}$ of those four $\langle|{\bm \omega}|(t)\rangle_s$ and $\langle u_zT(t)\rangle_s$  which pass through the two vertical diagonal planes in the cube. Thus, we now have two time series for each of both quantities. When the long-lived LSC are aligned along the diagonals (representing either $S_{7\pi/4}$ or $S_{3\pi/4}$ along one diagonal plane or the $S_{5\pi/4}$ or $S_{\pi/4}$ aligned along the other one), the heat flux along that diagonal plane maximizes. In a destabilization phase of the LSC, the vortices residing in the plane perpendicular to the LSC plane become intense and the corresponding vorticity magnitude time series peaks. Simultaneously, the heat flux along the direction of the LSC drops. This is shown in Fig.~\ref{trans_mech} for a shorter time window of $1000t_f$. The system undergoes a transition from the S$_{5\pi/4}$ to the S$_{7\pi/4}$ state. It is crucial to note that the transition is not smooth. The system toggles between the S$_{5\pi/4}$ state and the short-lived (S$_{n\pi/2}$ or decoherent state) before transitioning to S$_{7\pi/4}$ state. Therefore, Fig.~\ref{trans_mech} captures many of such destabilizing events. One such destabilizing events is observed in the rectangular box, of time length $100t_f$ free-fall times, in Fig.~\ref{trans_mech}, where the system transitions from long-lived S$_{5\pi/4}$ to short-lived S$_{n\pi/2}$ state. It is clearly observable that the heat flux drops along the LSC alignment when the vorticity magnitude peaks perpendicular to the LSC alignment. Within this time window, the Pearson correlation coefficient was obtained to be~$-0.71$. We computed the Pearson coefficient between the two signals for many such destabilizing events and always obtained a negative correlation coefficient (not shown). Away from the transition events, where the long-lived LSCs are stable, the Pearson coefficients are found to be either positive or slightly negative. This establishes our proposition that the intense vortical structures sitting perpendicular to the direction of LSC circulations triggers the destabilizing mechanism, whereby the long-lived LSC destabilizes and transitions to another long-lived LSC state via the short-lived or decoherent states.

\section{Data Reduction by Transition Manifold Framework}
\label{sec:methods}

The transition manifold framework (TMF) identifies low-- but possibly multi-dimensional collective variables, i.e., observables of the full state space that characterize essential dynamical phenomena. The present algorithm is unsupervised and data-driven, i.e., it does not require any physical knowledge about the turbulent flow. While the framework was designed originally for stochastic processes, we will demonstrate how it is applied to chaotic (here in a common, rather than a strict mathematical sense), but deterministic systems, such as the macroscopic RBC flow. We have already seen that the flow acts as a ``memoryless'' stochastic process on the largest scales, cf.\ the exponential distribution of persistence times. This suggests that there is a coarse-graining level in between on which the system can be well described as a stochastic process, making it amenable to the present framework. 

In the following, the variable $x$ stands for a state of the convection flow, i.e., one point in the high-dimensional state space $\mathbb{X}$, and not for a position vector as in previous sections~$\mathbf{x}\in \mathbb{R}^3$ stood for. 
\begin{figure*}
    \centering
    \includegraphics[width=\textwidth]{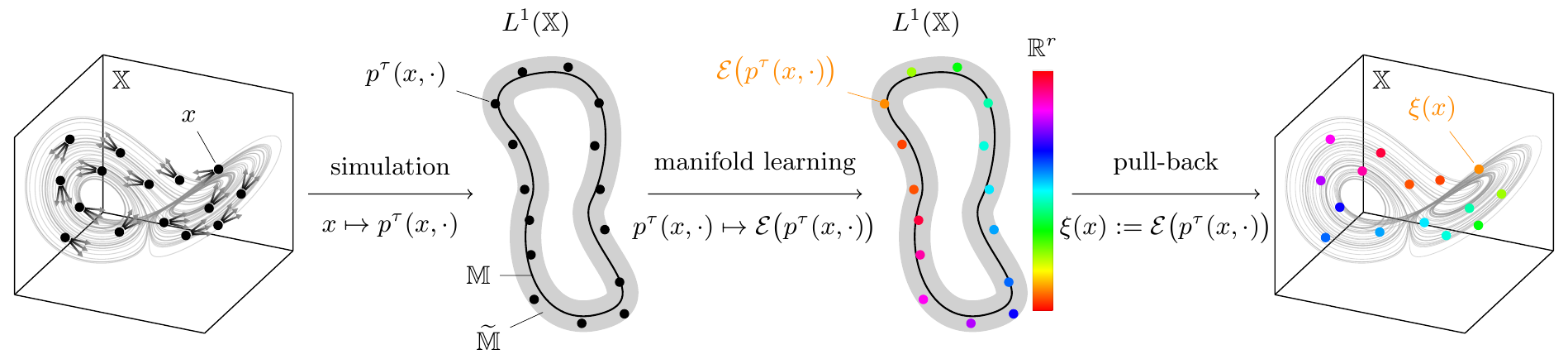}\\
    \hspace*{0mm} \textbf{a} \hspace{45mm} \textbf{b} \hspace{40mm} \textbf{c} \hspace{45mm} \textbf{d}
    \caption{Illustration of the transition manifold framework for a noisy system. (a) State space with initial states~$x$. The multiple arrows out of the dots in the picture underline the probabilistic nature of the system. (b) Initial states are mapped by their transition densities to $\smash{ \widetilde{\mathbb{M}} \subset L^1 }$, where they might accumulate around a low-dimensional manifold~$\mathbb{M}$. (c) A low-dimensional parametrization of $\smash{ \widetilde{\mathbb{M}} }$ is found, indicated by the colors. (d) This parametrization is trivially pulled back to the set of initial states to define the collective variable~$\xi$.}
\label{fig:TM_workflow}
\end{figure*}

\subsection{Original framework for stochastic systems}

We now line out in brief some basic mathematical relations to keep the manuscript self-contained. Consider a time- and space-continuous, homogeneous Markov process $\{X_t\}_{t\geq 0}$ (or short $\{X_t\}$) on a bounded state space $\X\subset \R^N$ (with variable initial condition~$X_0$) and an underlying probability space~$(\Omega,\mathcal{F},\operatorname{Prob})$. Assume that $\{X_t\}$ is ergodic and possesses a unique invariant density $\pi:\X\rightarrow \R^+$. We will denote by $\pi$ also the probability measure induced by $\pi$, and the Lebesgue measure by~$\mathrm{Leb}$. Under mild assumptions on the regularity of $\{X_t\}$, the statistics of the process are characterized by the family of \emph{transition density functions} $\{p^t\}_{t \geq 0} \subset L^1_{\pi \times \mathrm{Leb}}(\X \times \X)$, in the sense that 
\begin{equation}
\operatorname{Prob}[\,X_t\in A \mid X_0=x\,] = \int_A p^t(x,y)\ts dy.
\end{equation}
In other words, $p^t(x,\cdot)$ is the distribution of $X_t$ with starting distribution~$X_0\sim~\delta_x$. Written as a conditional distribution,~$(X_t \mid X_0=x) \sim p^t(x,\cdot)$.

Consider now some candidate collective variable $\xi:\X\rightarrow \R^r$, $r\ll N$, and let $\Y:=\xi(\X)$. As our interest in the RBC flow focuses on the most slowly evolving LSC phenomena and sub-processes of the system, we may informally call $\xi$ a ``good'' collective variable if a projection onto $\xi$ preserves the long-term statistics of the system. That is, for $t$ larger than some threshold timescale $\tau$, the statistics of $\{ \xi(X_t) \}$ will mimic that of $\{X_t\}$ in a certain sense. In the formal setting introduced above, this translates to 
\begin{equation}
\label{eq:CV_characterization}
\| p^t(x,\cdot) - \bar{p}^t(\xi(x),\cdot) \|_{L^1} \leq \varepsilon \quad \forall x\in\X, t\geq \tau,
\end{equation}
for some threshold lag time $\tau>0$, some family of functions $\{\bar{p}^t(\cdot,\cdot)\}_{t\geq 0} \subset L^1_{\bar{\pi} \times \mathrm{Leb}}(\Y \times \X)$ that are smooth in their first coordinate, and some small~$\varepsilon > 0$. Here, $\bar{\pi}$ denotes the pushforward of the measure $\pi$ by~$\xi$. We call a collective variable that satisfies \eqref{eq:CV_characterization} \emph{$\varepsilon$-consistent}.

Observe that eq.~\eqref{eq:CV_characterization} implies that the set of transition densities 
\begin{equation}
\label{eq:Mtilde}
\widetilde{\M} := \{p^\tau(x,\cdot)~|~x\in\X\} \subset L^1(\X)
\end{equation}
accumulates $\varepsilon$-closely around the $r$-dimensional manifold~$\M\subset L^1(\X)$,
\begin{equation}
\M :=\{ \bar{p}^\tau(z,\cdot)~|~z \in\Y\},
\end{equation}
the \emph{transition manifold}.
We emphasize that the transition manifold does not live in state space, and hence one should not think of it as a low-dimensional manifold connecting certain parts of state space, as heteroclinic orbits or manifolds would do.

Inversely, one can construct $\varepsilon$-consistent collective variables from parametrizations of the transition manifold. Let for this $\mathcal{E}:\M\rightarrow \R^r$ be any parametrization of $\M$, i.e., a one-to-one map between $\M$ and its image, and $\mathcal{Q}:\widetilde{\M} \rightarrow \M$ any map with the property
\begin{equation}
\left\| p^\tau(x,\cdot) - \mathcal{Q}\big(p^\tau(x,\cdot)\big)\right\|_{L^1} \leq \varepsilon.
\end{equation}
Under the assumption of~\eqref{eq:CV_characterization}, such a map exists and it can be shown that the collective variable
\begin{equation}
\label{eq:TMCV}
\xi(x) := \mathcal{E}\left( \mathcal{Q}\left(p^\tau(x,\cdot \right)\right)
\end{equation}
satisfies indeed~\eqref{eq:CV_characterization}. We call $\xi$ the \emph{transition manifold collective variable}. As in the case of the transition manifold, $\xi$ does not necessarily parametrize the time-evolution of every trajectory in state space, rather it parametrizes the \emph{progress of trajectories} during transitory dynamics in a collective manner.

As shown in \cite{BitEtAl17}, the process $\{ \xi(X_t) \}_{t\ge 0}$ admits dominant timescales (decay rates of correlations) which are $\mathcal{O}(\varepsilon)$-close to those of the original process~$\{X_t\}_{t\ge 0}$, if this is (stochastically) reversible. This quantifies how ``good'' a collective variable $\xi$ is, and paves the way for a constructive method that finds such collective variables of the transition manifold; cf.~Fig.~\ref{fig:TM_workflow}. A detailed discussion on the algorithmic realization of collective-variable computation can be found in appendix~\ref{sec:TMF_Pointwise}.

\subsection{Set-based approximation of collective variables for deterministic systems}
\label{sec:TMF_Voronoi}

For a deterministic flow with a map $\Phi^{\tau}$, the transition ``densities'' $p^{\tau}(x,\cdot)$ are Dirac distributions $\smash{ \delta_{\Phi^{\tau}(x)}(\cdot) }$ and hence not in~$L^1(\X)$. Yet, complex chaotic systems behave in many aspects as genuinely stochastic systems and their analysis is of probabilistic or statistical nature. This motivates our application of the TMF on deterministic chaotic systems such as the present one.

A solution to the above problem is offered by the concept of \emph{small random perturbations}~\cite{Kif86}. Intuitively, a family $(p_{\varepsilon})_{\varepsilon>0}$ of transition densities (or measures, in general) is a small random perturbation of the map $\Phi$ if in the space of measures, $\smash{ p_{\varepsilon}(x,\cdot) \stackrel{w}{\to} \delta_{\Phi(x)} }$ in the weak sense as~$\varepsilon\to 0$. Such small random perturbations retain certain dynamical properties of the original system in the limit of vanishing perturbation~\cite{Kha63,Kif86}. This is a kind of structural stability that we now assume extends to collective variables as well. The idea is to use small random perturbations with transition densities in~$L^1(\X)$ to transfer the TMF to deterministic systems.

The following discretization, that we propose to use in the collective variable computation pipeline for a deterministic flow $\Phi^{\tau}$, can be shown to provide a small random perturbation of the original system, with a transition density~\cite{Fr95}. More precisely, instead of computing the transition manifold collective variable $\xi$ at a single point, we will evaluate the average of $\xi$ over partition elements $\{A_1,\ldots,A_L\}$ of the state space~$\X$:
\begin{equation}
\xi(x) \rightsquigarrow \Xi(A) := \mathbb{E}[\, \xi(X)\mid X\sim\pi\vert_A \,].  
\end{equation}
This will give us (an approximation of) $\xi$ on the whole state space instead of just in isolated points, and at the same time allow its computation from long serial trajectory data, as opposed to simulation ``bursts''; cf. Algorithm~\ref{algo:Voronoi_CV}.

\begin{algorithm}[htb]
\caption{Set-wise computation of the transition manifold collective variables}
\label{algo:PointwiseRC}
\begin{algorithmic}[1]
\Require  Data set pairs $\mathcal{X}=\{x_1,\ldots,x_{N_s}\}$, $\mathcal{Y}=\{y_i,\ldots,y_{N_s}\}$ with $y_i = \Phi^\tau x_i$.

\State  Choose a partition $\{A_1,\ldots,A_L\}$ of $\X$ so that the metastable and transition regions are covered evenly.

\State Sort the data set $\mathcal{X}$ into the partition elements, i.e., for each $j$, compute the index set
$$
\mathcal{I}_j := \{ i~|~x_i\in A_j \}.
$$
The set
$$
\mathcal{Y}_j = \{y_i~|~i\in \mathcal{I}_j\}
$$
is then an empirical approximation of $p^\tau(A_j,\cdot)$.
\State Approximate the distance matrix $D\in \R^{L\times L}$,
$$
D_{ij} = d\left(p^\tau(A_i,\cdot), p^\tau(A_j,\cdot)\right)
$$
from the sampled $p^\tau(A,\cdot)$.
\State Apply an unsupervised, distance-based manifold learning algorithm to $D$.
\Ensure Approximation to transition manifold collective variable $\xi$, averaged over the sets $\{A_1,\ldots,A_L\}$, i.e. 
$$
\{\Xi(A_1),\ldots,\Xi(A_L)\}.
$$

\end{algorithmic}
\label{algo:Voronoi_CV}
\end{algorithm}
In our setting the data set pairs $\mathcal{X}, \mathcal{Y}$ will be constructed from a single long trajectory which is given by
\begin{equation}
\mathcal{T}:=\{x_0, \Phi^\tau x_0, \Phi^{2\tau} x_0,\ldots, \Phi^{N_s\tau} x_0\}
\end{equation}
where $\Phi^\tau$ represents the flow map of the dynamical system and $N_s$ is the number of snapshots, here $N_s=10^5$. We use
\begin{align}
\mathcal{X} &= \{x_0,\ldots, \Phi^{N_s-1}x_0\}\,,\\
\mathcal{Y} &= \{\Phi^\tau x_0,\ldots, \Phi^{N_s\tau}x_0\}\,,
\end{align}
where we use the lagtime~$\tau=1 t_f$. A trajectory output at this sampling rate is high enough to resolve the turnover process of fluid parcels in a LSC circulation roll.  

To obtain a partition of $\X$, we use a distance-based method. As the results of this method can strongly depend on the distance used, especially in very high-dimensional normed spaces, we will perform the partition on the trajectory $\mathcal{T}$ mapped to $z_i = \psi(x_i)\in\R^d$ in a lower-dimensional space by an auxiliary observation function~$\psi$. The viability of this and the choice of $\psi$ is discussed in the next section. The partition of $\mathcal{T}$ is then induced by the partition found for~$\psi(\mathcal{T})$. To partition the set $\smash{ z_1,\ldots, z_{N_s} \subset \R^d }$, we use a {\em Voronoi tessellation} of $\R^d$ with center points $\{z^{(1)},\ldots,z^{(L)} \}$ selected according to some sensible rule. A Voronoi tessellation of $\X$, with center points $\{x^{(1)},\ldots,x^{(L)} \}$, is a collection of sets $A_1,\ldots,A_L$ with $\smash{ \X = \cup_{i=1}^L A_i }$, such that $\smash{ x\in A_i \Leftrightarrow \|x-x^{(i)}\| \le \|x-x^{(j)}\| }$ for~$j\neq i$. In other words, the sets $A_i$ are formed by the points that are closer to $x^{(i)}$ than to any other center points~$x^{(j)}$. We here applied the $k$-means clustering algorithm with $k=L$ to $\psi(\mathcal{T})$ and use the resulting cluster centroids as Voronoi centers, as proposed in~\cite{bittracherDatadrivenComputationMolecular2018}. This has been demonstrated~\cite{bittracher_2021} to provide an even covering of the data points with center points. Further details are deferred to appendix~\ref{sec:TMF_Pointwise}.

\section{Collective variables for Rayleigh--B\'{e}nard flow}
\label{sec:results}

\subsection{Preprocessing by time-lagged independent component analysis (TICA)}

A flow state $x\in\X$ of RBC is originally infinite-dimensional, and even after the spatial discretization that is used for the direct numerical simulation it has millions of dimensions. The subsequent down-sampling from the original computational mesh still leaves us with a 12288-dimensional data space in the present cases, see also the next subsection. Representing general distributions in so many dimensions is a challenging task and it has proven itself useful to further pre-process the data by some coarse dimension reduction technique. Note that the assumption underlying the TMF is that the set $\widetilde{\M}$ in \eqref{eq:Mtilde} is almost a low-dimensional smooth manifold. This means, that almost every linear projection onto a sufficient but low-dimensional space is preserving the manifold structure of its core~$\mathbb{M}.$ \footnote{We note that this ``projection theorem'' is a useful quantitative statement, that is also used to prove Whitney's (weak) embedding theorem.}
\begin{figure}
\centering
\includegraphics[width=0.45\textwidth]{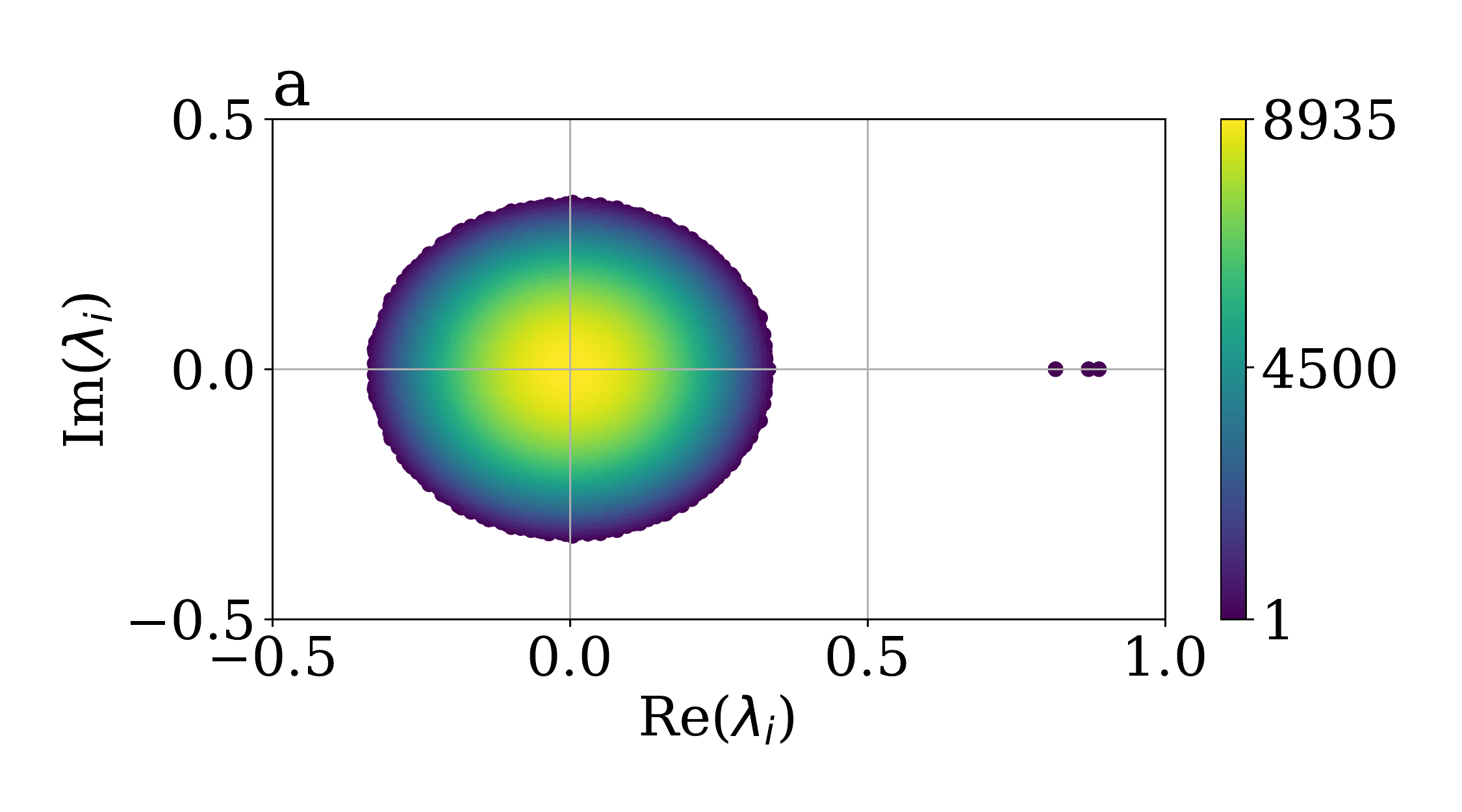}\\
\includegraphics[width=0.45\textwidth]{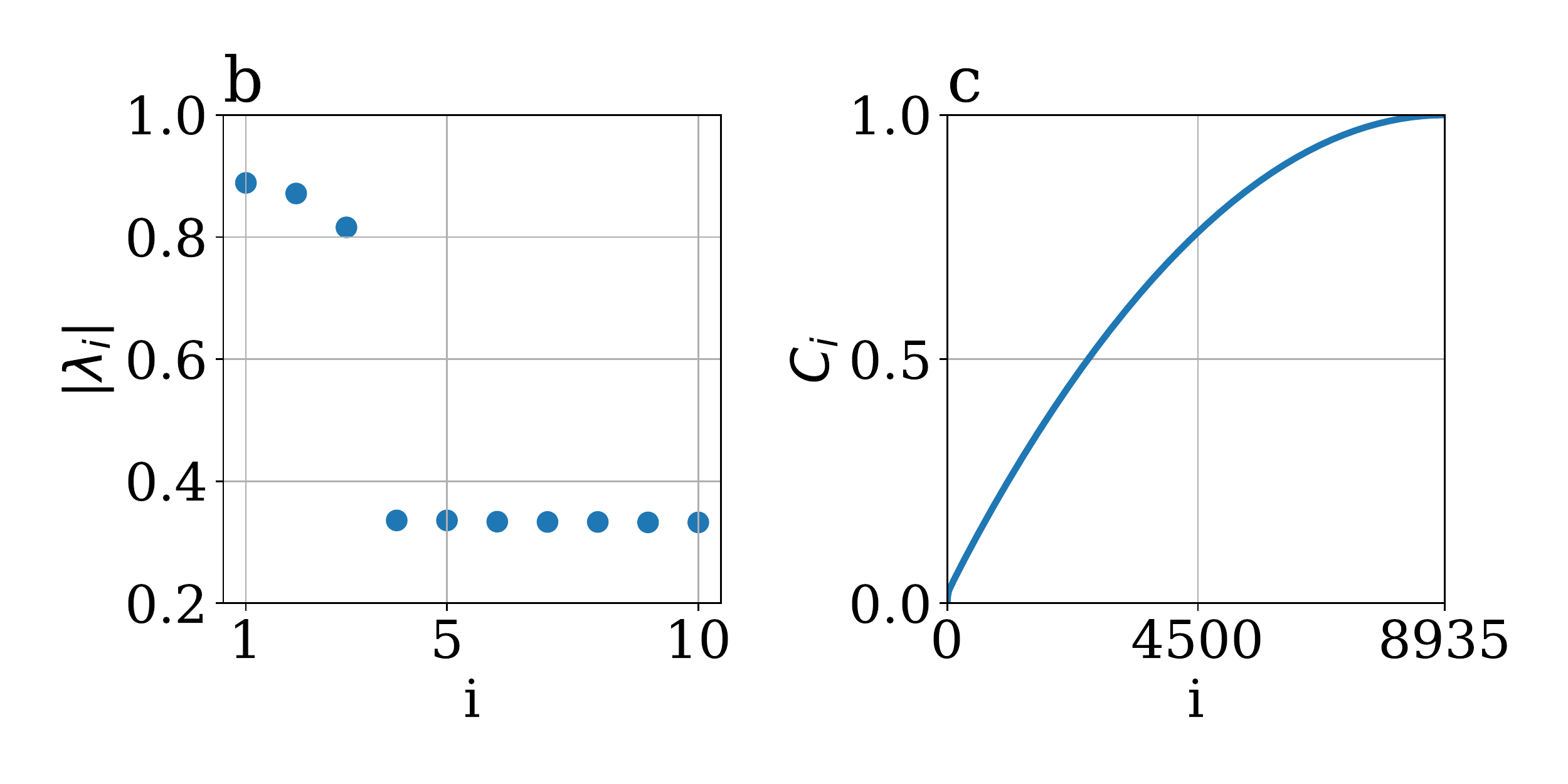}\\
\includegraphics[width=0.45\textwidth]{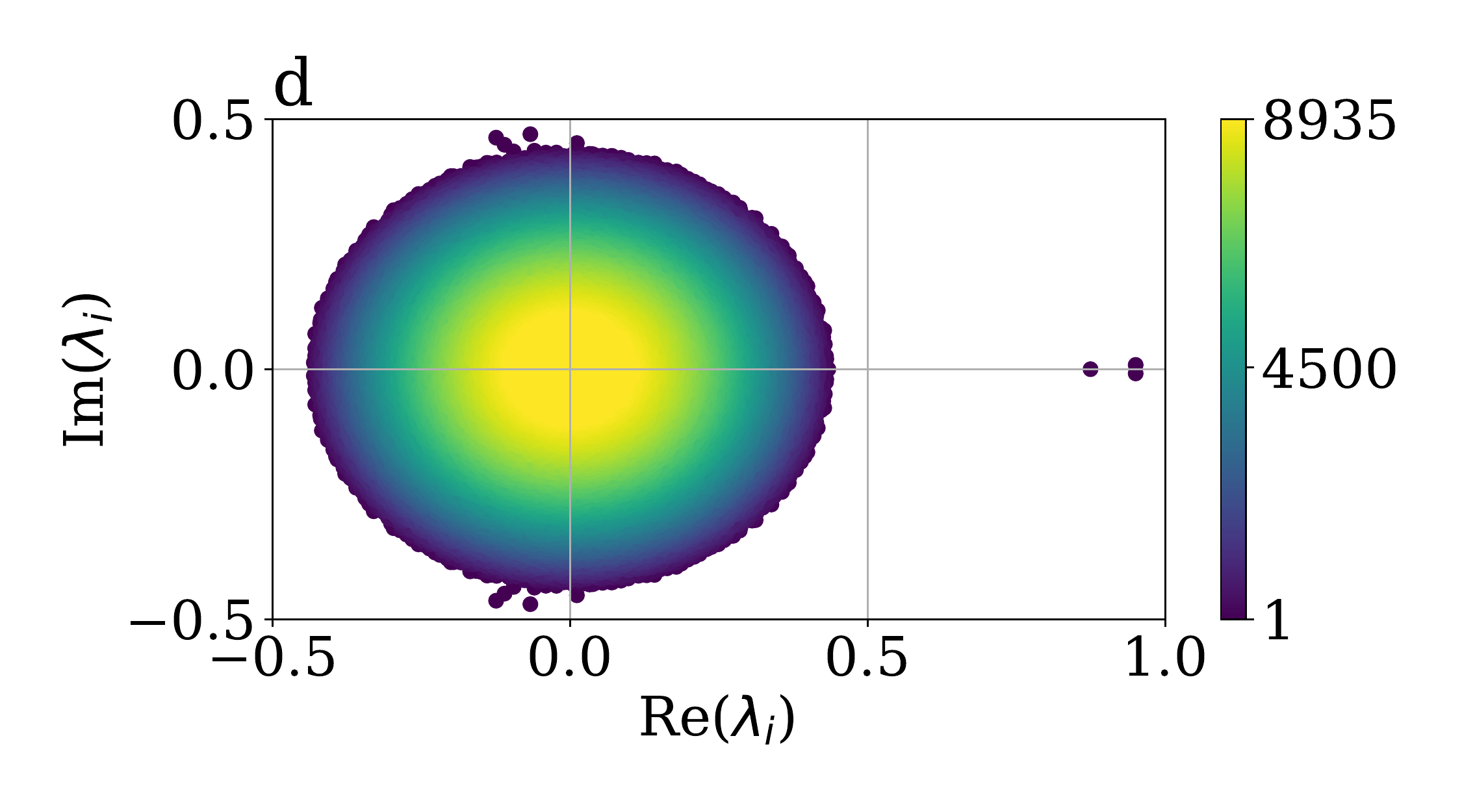}\\
\includegraphics[width=0.45\textwidth]{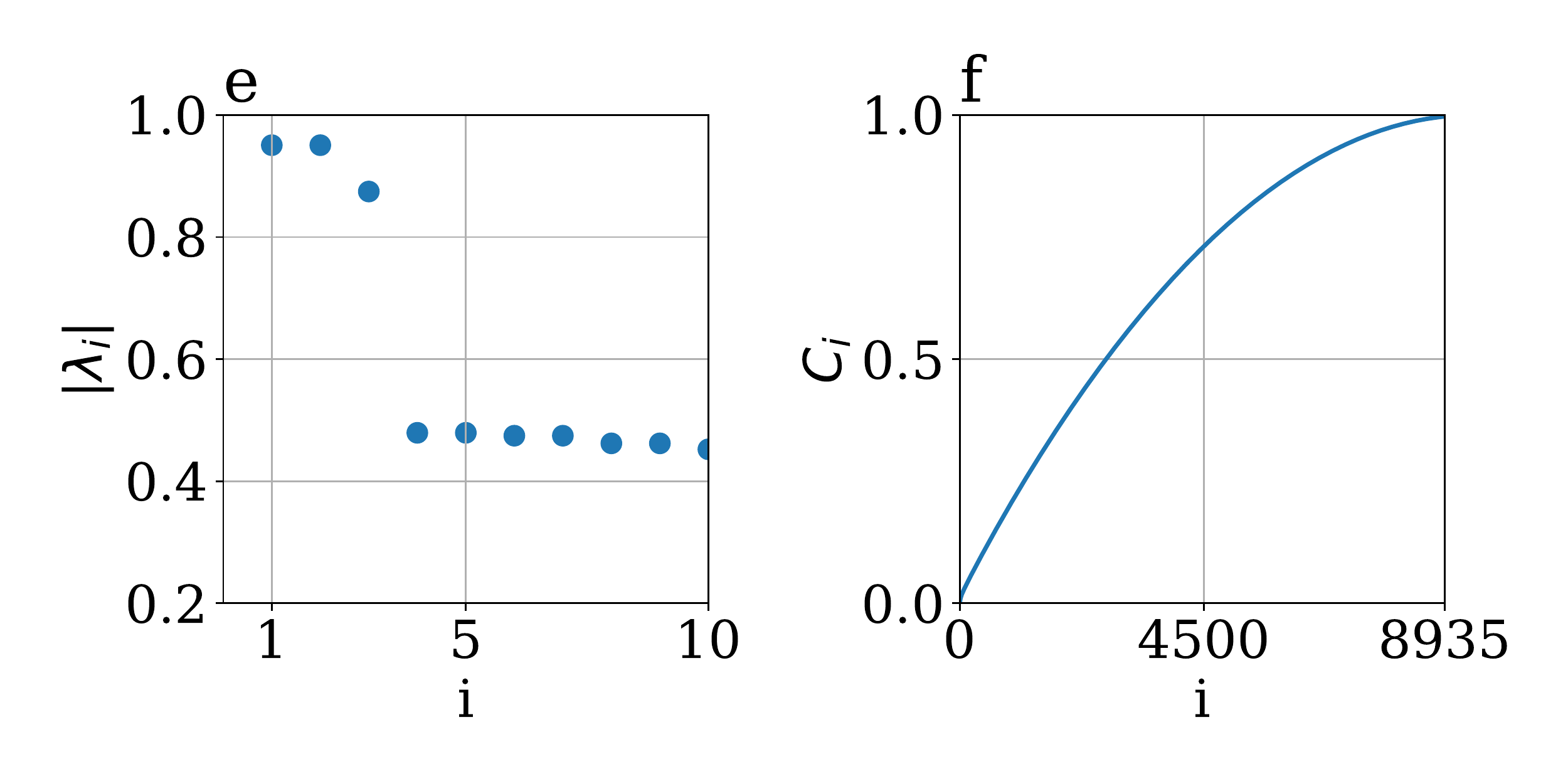}\\
\caption{(a) Real and imaginary part of the TICA eigenvalues for the run at $Ra=10^6$ (a--c) and $Ra=10^7$ (d--f). (a,d) The coloring corresponds to the eigenvalue index. (b,e) Absolute value of the leading 10 TICA eigenvalues. A spectral gap is clearly visible. (c,f) Cumulative kinetic variance $c_i$, indicating the portion of kinetic variance preserved by projecting onto the first $i$ TICA coordinates.}
\label{fig:TICA_eigenvalues}
\end{figure}

We will consider the pushforward of the distributions $p^{\tau}(x,\cdot)$, $x\in\X$, through a linear function~$\psi:\X\to \mathbb{R}^d$ with moderate range dimension~$d$. More precisely, as $(X_{\tau}\mid X_0=x) \sim p^{\tau}(x,\cdot)$, we will consider the distribution of $(\psi(X_{\tau}) \mid X_0=x)$ on~$\mathbb{R}^d$. Additionally to the above, this has the advantage that sampling this latter distribution is more accurate with the same amount of samples than~$p^{\tau}(x,\cdot)$, as~$d \ll N$.

The function~$\psi$ we choose to be composed by \emph{time-lagged independent component analysis} (TICA). In brief, TICA determines linear coordinates of the state that are ordered according to their so-called ``kinetic variances''; see appendix~\ref{sec:TICA} for further details on this data analysis algorithm. In this way, dominant TICA coordinates are expected to capture a lot of the dynamic variability and thus suggest themselves to be good observables to resolve the transition densities. Note that they are not expected to be able to parametrize the transitory dynamics---this would indeed be a lot to ask from linear observables for a highly nonlinear (turbulent) system.
\begin{figure*}
\centering
\includegraphics[width=0.85\textwidth]{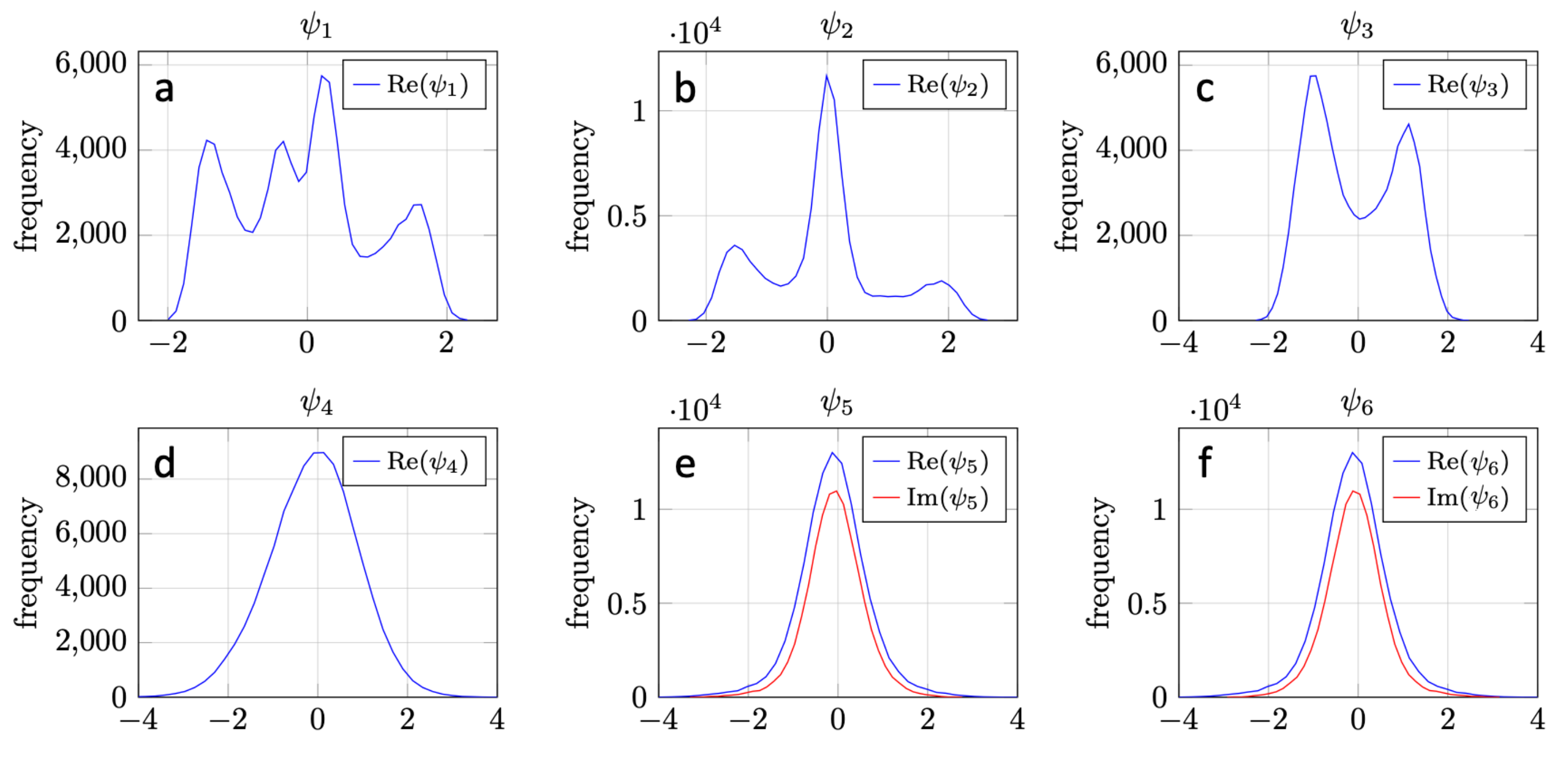}
\caption{Histograms of the values of the leading six TICA coordinates over the trajectory. The leading four coordinates are purely real, so only the real parts are plotted in panels (a--d); for the 5th and 6th leading coordinate in panels (e,f), the real and imaginary part is plotted. We observe a multimodal distribution in $\psi_1$ to $\psi_3$, and approximately normal distributions (real or complex) in the higher coordinates. There exist higher purely real TICA coordinates, which again are approximately normal distributed (not shown). Data are for $Ra=10^6$.}
\label{fig:TICA_eigenvectors}
\end{figure*}
\begin{figure*}
\centering
\includegraphics[width=0.9\textwidth]{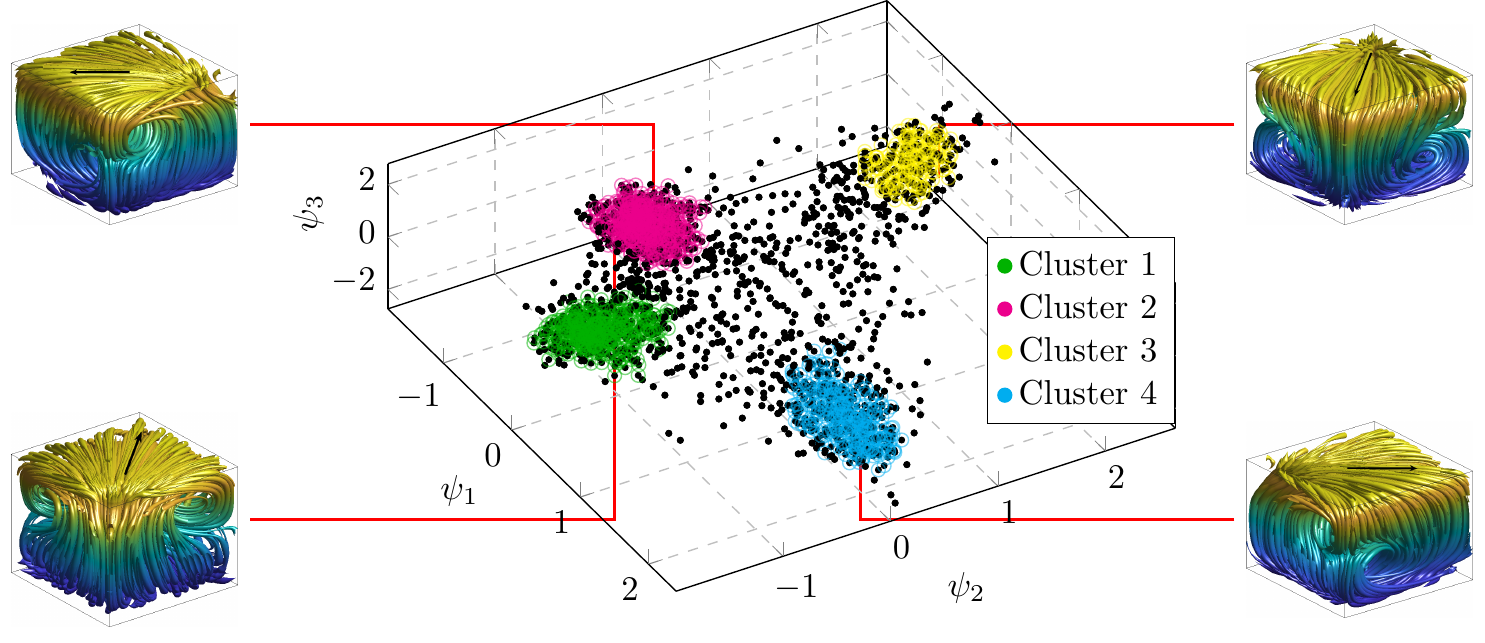}
\caption{Large-scale circulation states in the hyperplane which is spanned by the first three coordinates~$\psi_k$ determined by Time-lagged Independent Component Analysis (TICA). The center of the figure shows a subsample of the trajectory ($2000$ points) projected onto $(\psi_1,\psi_2,\psi_3)$-space. The points marked in color indicate the affiliation to four clusters as identified by the Density-Based Spatial Clustering of Applications with Noise (DBSCAN) algorithm in this projection. At the sides time-averaged velocity fields in the form of streamline plots are displayed which correspond to the four clusters. Data are for $Ra=10^6$.}
\label{fig:DBSCAN}
\end{figure*}

\subsection{TICA coordinates for Rayleigh--B\'{e}nard case}

Since we are interested in macroscopic and mesoscopic dynamical effects at the larger scales of the convective flow, we reduce the data amount and downsample from the original grid resolution of the DNS. The three components of the velocity vector field are therefore spectrally interpolated on a coarser uniform mesh with $16 \times 16 \times 16$ points. Hence, the RBC flow trajectory is embedded in a phase space with dimension~$N = 16^3 \times 3 = 12288$.

TICA is then run on this $N$-dimensional trajectory $(x_i)_{i=1}^{N_s}$ of length~$N_s =10^5$, producing the \emph{TICA coordinates} $\smash{ \psi_1(x_i), \psi_2(x_i), \ldots, \psi_N(x_i) }$ of the flow states~$x_i$. The TICA spectra in Figs.~\ref{fig:TICA_eigenvalues}~(a,d)  exhibit a clear gap after the third eigenvalue which is shown in panels (b,e) of the same figure. This gap indicates twice as much kinetic variance in each of the first three TICA coordinates than in any of the others and thus suggests to use $\smash{ \psi = (\psi_1,\psi_2,\psi_3)^\intercal }$ as the observation function introduced in the previous section (note that~$d=3$), see panels (c,f) of Fig.~\ref{fig:TICA_eigenvalues}. As detailed in appendix~\ref{sec:RBC_choices_bandidth}, for the transition manifold analysis there is no information gain in increasing the number of TICA coordinates, so the three leading ones are sufficient for both cases at hand.

Before proceeding to an analysis that considers the transitory dynamics of the system, let us analyze the data $\{\psi(x_i)\}_{i=1}^{N_s}$ from a static point of view, i.e., without taking information on succession of data points into account. Figure~\ref{fig:TICA_eigenvectors} shows histograms of the leading six TICA coordinates, showing a multi-modal structure of~$\psi_1,\psi_2,\psi_3$ at $Ra=10^6$. This suggests that large-scale structures of the RBC convection could be identified as clusters in the data mapped into~$\psi$-space. Indeed, plotting a subset \footnote{Sampled equidistantly in time with step $50$, hence $2000$ points.} of the trajectory in $\psi$-space reveals four densely-populated regions as seen in Fig.~\ref{fig:DBSCAN}. Applying the standard Density-Based Spatial Clustering of Applications with Noise (DBSCAN)~\cite{ester1996density} with parameters $\varepsilon=0.12$ and $\mathrm{minPts}=15$, we identify four clusters, i.e., four sets of trajectory index sets, that are robust to moderate parameter changes in the DBSCAN algorithm. Pulling these four index sets back into physical space, and averaging the corresponding flow fields, we see that they correspond to the four LSC states; see the streamline plots in Fig.~\ref{fig:DBSCAN} on the sides. Results for $Ra=10^7$ are similar (not shown).

The TICA-based analysis of the Rayleigh--B\'enard flow so far was intrinsically linear. With the following transition manifold framework we will advance to a nonlinear reduction. As mentioned above, although the TMF can be applied to arbitrary systems, it has a quantitative performance guarantee only for reversible systems. The RBC is a deterministic system and hence cannot be reversible. The remedy is that both the results in~\cite{Maity_2021} and the realness of the dominant TICA spectrum in Fig.~\ref{fig:TICA_eigenvalues}(a) indicate reversibility of the large scale processes.

\subsection{Transition manifold analysis}

While a projection onto the dominant three TICA coordinates clearly allows for the identification of the LSC states, it does not provide insight into the transition pathways between them.  For this reason, we now apply the set-based transition manifold algorithm, see Sec.~\ref{sec:TMF_Voronoi}. We have again shifted the technical details of this framework into appendix~\ref{sec:TMF_Pointwise}. The algorithm is the one which was already summarized in Algorithm~\ref{algo:Voronoi_CV}. It is now applied to the trajectory projected onto the leading three TICA coordinates.

The first step consists of the construction of a Voronoi tesselation of $\X$ that in particular resolves the transition regions, or equivalently, the selection of points $z^{(1)},\ldots,z^{(L)}$ that will become the centers of the Voronoi cells. For the reasons detailed in appendices~\ref{subsec:state_space} and~\ref{subsec:transition_density}, we choose the $k$-means clustering algorithm with a cluster count of $k=L=2000$ for the center point selection. This results in an average number of 50 trajectory snapshots being designated to each Voronoi cell. In previous studies involving the TMF, comparable sample numbers have been demonstrated to capture the dynamical properties of interest~\cite{bittracherDatadrivenComputationMolecular2018,bittracher_2021}. The selected center points $z^{(1)},\ldots, z^{(L)}$ slightly undersample the cluster areas, and cover the transition areas more densely, when compared to the equidistant subsampling. Note that, by the nature of $k$-means, the center points are in general not actual states from the data set (i.e., the trajectory) but instead the centroids of certain subsets of the data set.

\begin{figure*}[!ht]
\centering
\includegraphics[width=1\textwidth]{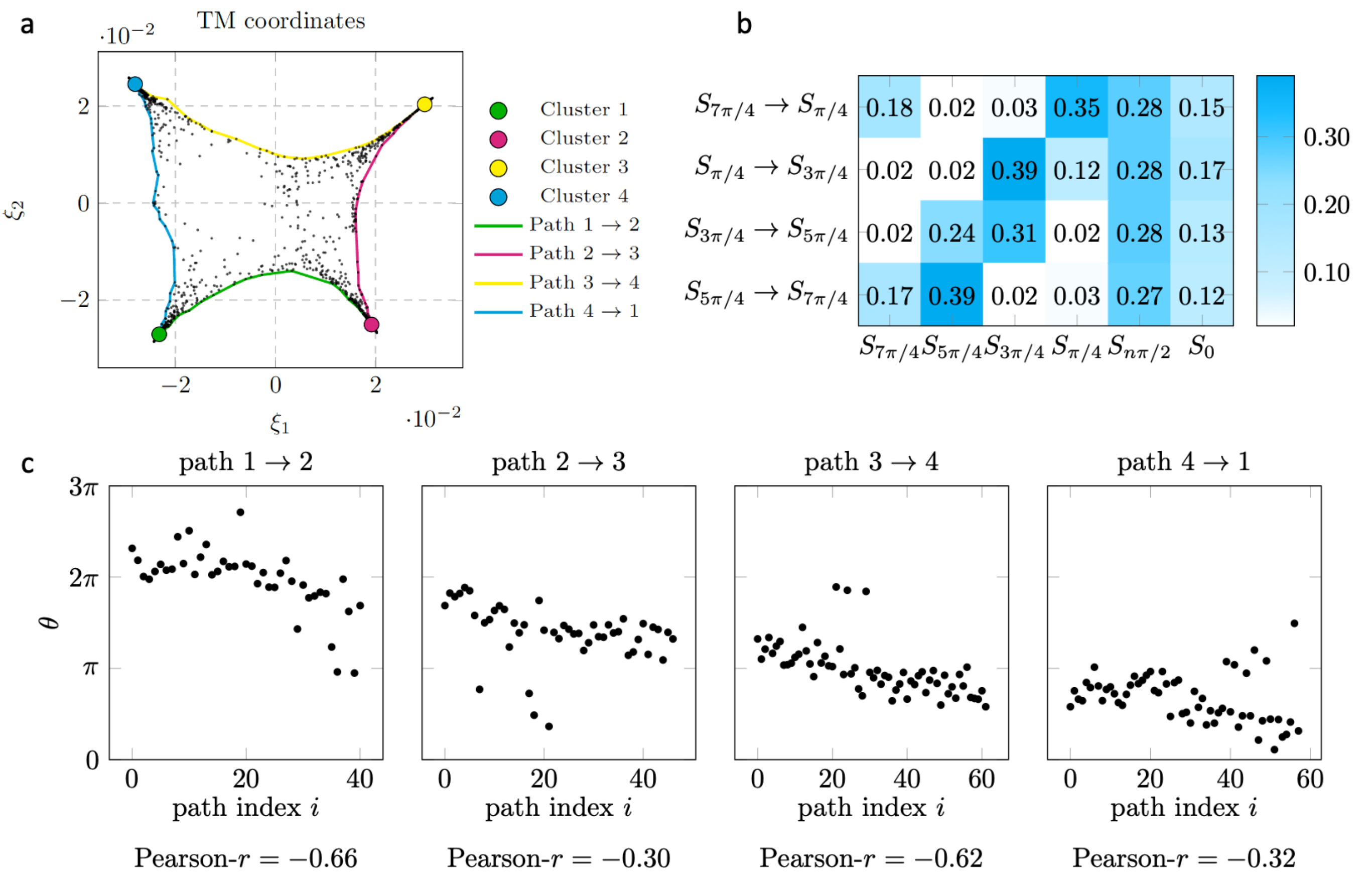}
\caption{Transition manifold analysis for the Rayleigh--B\'{e}nard flow. (a) Shortest pathways between the cluster centers which were determined in the transition manifold space spanned by the two-dimensional collective variable~$\xi$. (b) Portion of time the transition path spends in or between each of the six large-scale circulation states. (c) Comparison of the progress along the individual transition pathways, listed in the form of the indices of the points along the paths, to the midplane angle~$\theta$. One visually observes a strong linear correlation, which is also confirmed by a high Pearson correlation coefficient. Note that, due to the periodic nature of $\theta$, plotting its raw value would not convey its trend along the paths correctly. Hence, in order to obtain these results, a preprocessing step has been conducted to possibly shift each $\theta_i$ by $\pm 2\pi$, according to which value gives the closest distance to~$\theta_{i-1}$. Data are for $Ra=10^6$.}
\label{fig:4}
\end{figure*}

Next we apply Algorithm~\ref{algo:Voronoi_CV} with the Voronoi cells $A_1,\ldots, A_L$ associated to the center points~$z^{(1)},\ldots, z^{(L)}$. For the distance measure between the transition densities we use Maximum Mean Discrepancy (see appendix~\ref{subsec:MMD}) and we use diffusion maps to learn the collective variable $\xi$ from the resulting distances, see appendix~\ref{subsec: manifold learning}. The resulting analysis, see appendices~\ref{sec:bandwidth} and~\ref{sec:RBC_choices_bandidth}, shows that the associated transition manifold is at most two-dimensional, hence we obtain a two-dimensional collective variable:~$\xi(x_i) \in \R^2$. The collective variable $\xi$ of the Voronoi centers is shown in Fig.~\ref{fig:4}(a) as black dots.
The data now accumulates in the four corners of this collective-variable space, corresponding to the LSCs. 

In contrast to the TICA embedding in Fig.~\ref{fig:DBSCAN}, however, the clusters are much more concentrated. This indicates that, while two points $x_i,x_j$ from the same cluster may be considerably distinct in physical or TICA-$\psi$-space, their \emph{dynamical} distance, i.e., the Maximum Mean Discrepancy between the densities $p^\tau(x_i,\cdot)$ and $p^\tau(x_j,\cdot)$ is much smaller. This allows us to better group states by their future statistics, i.e., \emph{identify states that have the same long-term evolution}.

\subsection{Transitory dynamics in collective variables}

As close-by points on the transition manifold have similar evolution, the shortest path in collective-variable space is likely to have distinctive dynamical relevance. To validate this, we consider shortest paths between the corners in Fig.~\ref{fig:4}(a); these are the colored curves. More precisely, for two cluster centers $c_1,c_2\in \mathbb{R}^2$, we consider the interconnecting line 
$$
\gamma(r) := rc_1 + (1-r)c_2,\qquad r\in[0,1],
$$
and define as the discrete pathway from $c_1$ to $c_2$ between sample points of the transition manifold as
$$
\Gamma(r) := \operatorname{argmin}_{i=1,\ldots,L} \big\| \gamma(r) - \xi(z^{(i)})\big\|.
$$
Note that these paths are not trajectories of the system. Progress along the individual pathways shows strong correlation to the midplane angle $\theta$ (recall Fig.~\ref{fig:1}), as indicated in Fig.~\ref{fig:4}~(c). The imperfect correlation between $\theta$ and $\Gamma(r)$ is expected as the classification of circulation states through the midplane angle. It is based on (i) a geometric-physical intuition and (ii) on solely a time-instantaneous flow-field information, in contrast with dynamical information as in the TMF. Yet, there is a clear correlation between the results of these two classification methods. These transition pathways align well with the heuristic classification from Sec.~\ref{subsec:detection_LSC}. This is further underlined by the portion of the pathways spent in the respective LSCs, shown in Fig.~\ref{fig:4}(b). We observe that a pathway connecting two corners associated with two long-lived LSCs spends most of the time in these states, while transitioning through the short-lived LSC and decoherent states and notably spending barely any time in other LL-LSCs.

\begin{figure*}[!ht]
\centering
\includegraphics[width=1\textwidth]{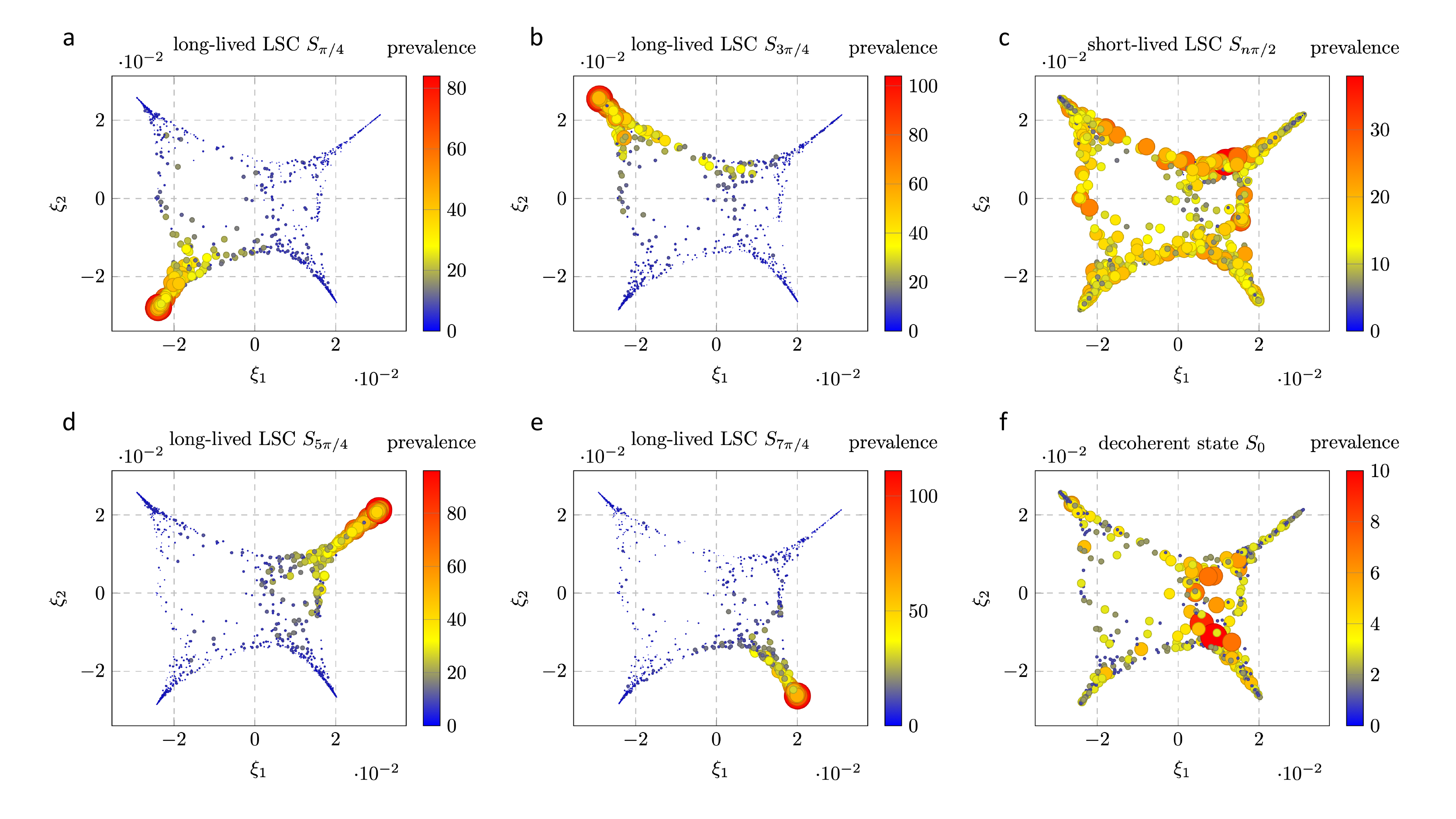}
\caption{Prevalence of the four long-lived (a,b,d,e), the short-lived (c), and the decoherent (f) states among all trajectory snapshots for the case of~$Ra=10^6$. Both the color and the size of a point indicate the number of trajectory snapshots that fall within the Voronoi cell corresponding to the point, and belongs to one of the six states, as identified by the method used in Fig.~\ref{fig:1} and described in detail in Sec.~\ref{subsec:detection_LSC}. Data are for~$Ra=10^6$.}
\label{fig:5}
\end{figure*}

Figure~\ref{fig:5} illustrates the prevalence, i.e., the frequency of appearance in associated Voronoi cells, of the different LSC states in collective-variable space. LSCs of trajectory snapshots are again computed by the classification method used in Fig.~\ref{fig:1}. We observe that snapshots of the long-lived LSCs each concentrate in one corner of the transition manifold. Moreover, the short-lived LSC tend to be evenly distributed along the edges and corners of the structure, whereas decoherent snapshots have a stronger tendency to fall in the center of the transition manifold. These results demonstrate that the geometry of the transition manifold encodes and highlights the long-term stability and transition dynamics of the system.

Finally, we note that the transition manifold analysis shows also that there are no flow reversals. For this, an alternative parametrization of the transition manifold---and thus slightly distorted collective variables---were considered in Fig.~\ref{fig:RBC_TICA03_dmaps}(a) in appendix~\ref{sec:RBC_choices_bandidth}, that emphasizes small-scale geometric features. This parametrization shows no pathways connecting diagonally opposite corners of the transition manifold, which correspond to long-lived LSCs of reverse circulation direction.

We have repeated the TMF analysis for the data at~$Ra=10^7$ (not shown). The pipeline of data processing is the same resulting first in 3 TICA coordinates as already shown in Fig.~\ref{fig:TICA_eigenvalues}(d--f). The TMF framework gives us again 4 distinct clusters. The MMD histogram, which is shown in Fig.~\ref{fig:kernelparameters} in appendix~\ref{subsec:MMD}, suggests that the data are more strongly clustered which can also be seen from Fig.~\ref{fig:1}(b). However, there is a difference to the case at~$Ra=10^6$. Namely, the structure is not quite two-dimensional anymore. It is rather three-dimensional, or maybe even more. The third dimension does however not give the transition manifold any additional interpretable structure, it only seems to ``thicken'' the manifold, i.e., as if it were merely noise. It can be expected that with increasing Rayleigh number, the dimension of the transition manifold will increase, an aspect that we have to leave for work in the future.

\section{Summary and discussion}
\label{sec:summary}

This work presented a data-driven description of the crossover dynamics from one large-scale state into another in a turbulent Rayleigh--B\'{e}nard convection flow -- a paradigm for a deterministic macroscopic and complex nonlinear dynamical system. The RBC flow in a closed cubic cell is known to rest in macro-states in the form of large-scale circulation rolls along the diagonals for hundreds of convective time units before switching rapidly into another diagonal flow state. The specifics of these transitions were in the focus of our study. They are characterized by a complex interplay of the thermal plumes which build up the large-scale circulation and fluctuating high-amplitude vortices at the sides of the LSC, as we discussed in sections~\ref{DNS} and~\ref{sec:LSC} in detail. See also Figs.~\ref{vor_dist} and~\ref{fig:1} which displayed the transition. This dynamics is fully three-dimensional and includes the majority of the degrees of freedom of our dynamical system at hand.   

We applied the transition manifold framework to the present system, which was originally developed for stochastic microscopic systems, such as for conformational transitions of macromolecules. In our macroscopic case, a significant reduction to 2 coordinates, denoted as collective variables, was achieved. The initial computational grid for the 4 turbulent fields $(u_x,u_y,u_z,T)$ with a total of at least $N_{\mathrm{full}}\gtrsim 3.5 \cdot 10^6$ degrees of freedom was first downsampled in both runs to $3\times 16^3 = 12288$ degrees of freedom for $(u_x,u_y,u_z)$. This forms the starting point of our reduced description of the transition dynamics. 

The subsequent data-driven calculation of the 2 coordinates that span the nonlinear transition manifold was pre-conditioned by a time-lagged independent component analysis (TICA) that reduces the 12288-dimensional data space. TICA determines linear coordinates which allowed us to identify LSC states as shown in Fig.~\ref{fig:DBSCAN}. On the TICA-preprocessed time series we then ran the transition manifold analysis to find that actually two coordinates that describe its transition dynamics. The projection of the dynamics on these collective variables improved the distinction of the clusters significantly which can be seen when comparing Figs.~\ref{fig:DBSCAN} and~\ref{fig:5}. The LSC states are clearly assigned with the cusps of the transition manifold. Furthermore, Fig.~\ref{fig:5} demonstrates clearly that the short-lived LSCs are lined up along the boundary of the manifold, connecting the cusps, and that the decoherent state is found in the interior; i.e., all six macrostates can be properly separated in the plane which is spanned by the two collective variables. The new collective variables also demonstrate that flow reversals are not observed for the chosen parameter sets, thus we can conclude that they are very rare in the present 3d RBC convection case.

In future work, one could use these reduced coordinates to identify a surrogate dynamics of the large-scale processes in the system. A further interesting extension of this framework can be obtained by the study of convection flow configurations in closed cells with a somewhat larger aspect ratio. Then multiple LSC rolls will fill the simulation domain. The first step would comprise a verification of the Markov property of the large-scale flow dynamics as done for the present case in ref.~\cite{Maity_2021}. It is expected that more long-lived macrostate configurations are possible then and that in turn more than 2 collective variables are required to describe the transitions within the TMF framework.  

\subsection{Data processing pipeline summary}
Several specifics of the applied methods are outlined in detail in the appendices including additional figures and algorithm tables. They contain technical details on the transition manifold framework and the intermediate reduction by time-lagged independent component analysis. Here, we briefly summarize the data processing pipeline of learning the collective variables once more in 5 subsequent steps. These are:

\begin{enumerate}

\item
Downsampling of the grid from $N_{\mathrm{full}}\approx 3.54 \times 10^6$ and $3.84\times 10^6$ degrees of freedom for $Ra=10^6$ and $10^7$, respectively, to~$N = 12888$ degrees of freedom. This is justified by our focus to processes that show up on mesoscopic scales and larger.

\item 
Computation of the time-lagged independent component analysis coordinates $\psi_1,\ldots, \psi_N$ and setting of  $\psi := (\psi_1,\psi_2,\ldots,\psi_d)^\intercal$ with~$d=3$ as observables of the high-dimensional original system. This reduction follows the assumption that the transition densities are low-dimensional and that density estimation is more accurate for~$d\ll N$. 

\item
Subsequent Voronoi tesselation of $\{ \psi(x_i) \mid i=1,\ldots,N_s \}$ to find Voronoi cells $A_i\subset\R^d$, $i=1,\ldots,L=2000$. This is justified by the fact that complex turbulent flows behave like stochastic systems.

Branching off from the main pipeline, we applied DBSCAN to the cell center points and associated the clusters with LSCs of the convection flow.
    
\item 
Approximate the $\psi$-pushforward of transition densities $p^{\tau}(x_i,\cdot)$ by the samples $\{ \psi(x_{j+\tau}) \mid \psi(x_j) \in A_i \}$ and use maximum mean discrepancy to compute the distance matrix $D\in\mathbb{R}^{L \times L}$ between them. The feasibility of this step was demonstrated in \cite{bittracher_2021}, where the maximum mean discrepancy turns out to allow for a ``quasi-embedding''.

\item
Compute transition manifold by applying diffusion maps to $D$ to obtain the estimated dimension two and the associated collective variable~$\xi = (\xi_1,\xi_2)^\intercal$.
\end{enumerate}

\acknowledgements 
The work of P.M. is supported by the grant nos. SCHU 1410/29-1 and SCHU 1410/30-1 of the Deutsche Forschungsgemeinschaft (DFG). A.B. is supported by DFG through grant CRC~1114 ``Scaling Cascades in Complex Systems'', Project Number 235221301, Project B03 ``Multilevel coarse graining of multiscale problems''. P.K. has been partially supported by the DFG through grant CRC 1114 ``Scaling Cascades in Complex Systems'', Project Number 235221301, Project A08 ``Characterization and prediction of quasi-stationary atmospheric states'' and under Germany’s Excellence Strategy—The Berlin Mathematics Research Center MATH+ (EXC-2046/1 project ID: 390685689). The comprehensive long-term simulations were conducted at the compute cluster of the University Computing Centre at Technische Universit\"at Ilmenau (Germany).

\appendix
\section{Time-lagged Independent Component Analysis (TICA)}
\label{sec:TICA}

The time-lagged independent component analysis (TICA) is applied to preprocess the simulation data. TICA provides a reduction of the data to a low-dimensional hyperplane in the state space that will be the starting point for the subsequent transition manifold analysis to determine the collective variables.

\subsection{TICA algorithm}
\label{subsec:TICA_algorithm}

The evolution of the LSCs proceed in a high-dimensional configuration space (or phase space) and consist of both slowly evolving or metastable LL-LSC states and fast-evolving transition states which includes SL-LSC and decoherent states. Due to the high dimensionality of the data it is demanding to extract the basic configuration and transition states of the system. This suggests employing dimension reduction techniques for efficacious handling of the generated data.

Perhaps the most widely used linear dimension-reduction technique is principal component analysis~(PCA). While PCA finds high-variance linear combinations of the input degrees of freedom, TICA~\cite{molgedey_TICA_1994} is a linear transformation method based on a variational approach, which transforms a set of high-dimensional input coordinates to a set of low-dimensional output coordinates having maximal autocorrelation. It can thus be seen as a dynamic version of PCA. TICA has been developed in signal processing~\cite{molgedey_TICA_1994} and more recently applied in molecular dynamics~\cite{perez_tica_2013,nuske_2014,schultze_2021}. Its use has two objectives, namely (i) to reduce the dimensionality of the configuration space for a faster processing of the data, and (ii) extract the slow order parameters which can give us an idea about the slowly evolving states.

Another method frequently applied in fluid dynamics is dynamic mode decomposition (DMD)~\cite{schmid2009dynamic,schmid2010dynamic}, where a linear model is fitted to the observed time series in a least squares sense. It should be noted that TICA and DMD are essentially equivalent on an algorithmic level, as shown in~\cite[sec.~3.2]{KlEtAl18}. However, while DMD is designed and used to visualize and forecast evolving fluid flows by a linear model, TICA is utilized to produce a set of (reduced linear) coordinates that explain most of the dynamical variability in the system. We shall thus employ the TICA-terminology in the following.

We begin with a $N$-dimensional dynamical system trajectory $\mathbf{x}(t)=(x_1(t),..., x_N(t))^\intercal \in \mathbb{R}^N$. Here, Cartesian coordinates are taken as the three components of the velocity vector field are spectrally interpolated on a uniform $16 \times 16 \times 16$ mesh (coarser than the computational one). Hence, the RBC flow trajectory is embedded in a phase space with dimension $N = 16^3 \times 3 = 12288$. Note that we do not include the temperature field in the analysis. The input data should obey a zero mean, thus we will use
\begin{equation}
\mathbf{x'}(t)=\mathbf{x}(t) - \langle \mathbf{x}(t) \rangle_t\,,
\label{mean_free_data}
\end{equation}
where, $\langle \cdot \rangle _t$ denotes time averaging. Thereafter, we compute the autocovariance matrix $C(\tau)$ at various lag times $\tau$ which is given by 
\begin{equation}
C_{ij} (\tau) = \langle x'_i(t)x'_j(t+\tau)\rangle_t \in \mathbb{R}^{N\times N}\,.
\label{cov_matrix}
\end{equation}
The matrix elements of $C$ are evaluated by
\begin{equation}
C_{ij} (\tau) = \frac{1}{N_s-\tau-1}  \sum_{t=1}^{N_s-\tau} x'_i(t) x'_j(t+\tau)\,,
\label{cov_matrix_calculation}
\end{equation}
where $t$ and $\tau$ are integer-valued multiples of the output time interval, $1 t_f$. Times are thus associated with integers which run from 1 to $N_s=10^5$ in the present case.  The TICA eigenvectors $\mathbf{\Psi}_k \in \mathbb{R}^N$ can be obtained as the solution of the generalized eigenvalue problem
\begin{equation}
C(\tau)\mathbf{\Psi}_k = \lambda_k C(0)\mathbf{\Psi}_k\,.
\label{tica_eigen}
\end{equation}
with the $k$-th TICA eigenvalue~$\lambda_k$. The \emph{TICA coordinates} (or TICA modes) $\psi_1(x'(i)), \psi_2(x'(i)),\ldots$ are then the coefficients of the $x'(i)$ represented in the basis composed of the TICA eigenvectors~$\mathbf{\Psi}_k$. We note that with the data matrices
\begin{equation}
\begin{aligned}
X &{:=} \begin{bmatrix}
\vert & & \vert \\
x'(1) & \cdots & x'(N_s-\tau)\\
\vert & & \vert
\end{bmatrix} \quad\text{and}
\\
Y &{:=} \begin{bmatrix}
\vert & & \vert \\
x'(1+\tau) & \cdots & x'(N_s)\\
\vert & & \vert
\end{bmatrix},
\end{aligned}
\end{equation}
one often writes 
\begin{equation}
\begin{aligned}
C(0) &= \frac{1}{N_s-\tau-1} XX^\intercal \quad\mbox{and}\\ C(\tau) &= \frac{1}{N_s-\tau-1} XY^\intercal\,.
\end{aligned}
\end{equation}

\subsection{Coordinate reduction}

We compute the TICA coordinates for a lag time of $\tau=50$ using the PyEMMA package~\cite{SchEtAl15}.  Figure~\ref{fig:TICA_eigenvalues}(a) shows the eigenvalues $\lambda_i,~i \geq 1$ of the autocovariance matrix which will be called \emph{TICA eigenvalues} in the following.

It is of separate interest, but we note that TICA (and DMD) provide approximations to the composition operator associated with the dynamics, named after Koopman, see~\cite{KlEtAl18} and references therein. As such, the TICA spectrum provides an admittedly coarse approximation of the system's spectrum, and hence of the decay of correlations in the dynamics.

We observe that the majority of the spectrum is confined to a complex disk of radius smaller than~1. This is a typical situation for quasi-compact transfer (and Koopman) operators, that arise for complicated dynamics exhibiting a certain degree of chaoticity or mixing~\cite[sections~1.3 and~2.3 in particular]{baladi2000positive}. Furthermore, the dominant (by absolute value) TICA eigenvalues are purely real, which suggests that on long time scales associated with these eigenvalues the system behaves in a statistically reversible manner. Indeed, transitions from a LSC to other ones do not seem to have a cyclic tendency. The absolute value of a TICA eigenvalue is also known as the \emph{kinetic variance} of its corresponding eigenvector, and can be an indicator for how much ``dynamical information'' is preserved by projecting onto this coordinate~\cite{noe_kinetic_2015}. The kinetic variances of the leading 20 coordinates are shown in Fig.~\ref{fig:TICA_eigenvalues}(b). The spectral gap after $\lambda_3$ is immediately apparent, and hints at the disproportional dynamical significance of the first three eigenvectors.

The \emph{cumulative kinetic variance} is now defined as the sum of the kinetic variances up to a certain TICA eigen pair, relative to the sum of all kinetic variances:
\begin{equation}
c_i :=\frac{\sum_{k=1}^i |\lambda_k|}{\sum_{k=1}^{N} |\lambda_k|}
\end{equation}
The course of $c_i$ in $i$ is shown in Fig.~\ref{fig:TICA_eigenvalues}~(c). We observe a flattening of the curve, indicating a greater impact of lower TICA coordinates (which is clear as the eigenvalues are ordered by decreasing absolute value), but not a visible jump or kink after~$c_3$. Moreover, the curve exhibits a considerably shallower slope than what one is accustomed to in the analysis of for example many molecular dynamical data sets~\cite{perez_tica_2013}. This tells us that a large portion of the kinetic variance is generated by medium- and small-scale processes and not overly concentrated in the global large-scale processes, which are well represented by the leading three TICA coordinates (see Fig~\ref{fig:DBSCAN} in the main text). Hence, while $\lambda_i$, $i={1,2,3}$, may contain a disproportional amount of information, a much greater number of TICA coordinates would be required to resolve the dynamics of the system by a linear model more or less completely. Also, note that in the single precision floating point arithmetic used by PyEMMA, $c_{8935}$ is equal to $1$, hence the data set can be reduced by $N-8935=3353$ dimensions without measurable loss of kinetic variance, where $N=12288$ is the original system dimension.

Figure~\ref{fig:TICA_eigenvectors} visualizes the leading six TICA coordinates by the histograms of their values along our trajectory. Corresponding to the three leading purely real eigenvalues, the leading three coordinates are also purely real. Moreover, the coordinates belonging to the well-separated, dominant part of the spectrum possess a multi-modal distribution, whereas the coordinates belonging to eigenvalues in the aforementioned complex disk are approximately normal-distributed.

The multi-modal structure of $\psi_{1,2,3}$ raises hope that it allows the identification of large-scale structures of the RBC convection. Indeed, plotting a subset \footnote{Sampled equidistantly in time with step $50$, hence $2000$ points.} of the trajectory in $(\psi_1,\psi_2,\psi_3)$-space reveals four densely-populated regions (see Fig.~\ref{fig:DBSCAN} in the main text). Applying the Density-Based Spatial Clustering of Applications with Noise (DBSCAN)~\cite{ester1996density} with parameters $\varepsilon=0.12$ and $\mathrm{minPts}=15$, we identify four clusters, i.e., four sets of trajectory index sets, that are robust to moderate parameter changes in the DBSCAN algorithm. Pulling these four index sets back into physical space, and averaging the corresponding flow fields, we see that they correspond to the four LSC states (see again Fig.~\ref{fig:DBSCAN}, on the sides).

\section{Algorithmic realization of the transition manifold framework for stochastic systems}
\label{sec:TMF_Pointwise}

In practice, neither the transition manifold $\M$, nor the maps $\mathcal{E}, \mathcal{Q}$ are known, and $p^\tau(x,\cdot)$ is only known empirically (e.g., by starting a large number of DNS runs with different random seeds in $x$ up to time $\tau$). The algorithmic strategy therefore consists of \emph{learning} the parametrization $\mathcal{E}$ and the projection $\mathcal{Q}$ simultaneously from a finite subsample of the set~$\smash{ \widetilde{\M} }$. We first give an overview of the algorithm and then explain the individual steps in more detail:

\begin{algorithm}[H]
\floatname{algorithm}{Algorithm}
\caption{Point-wise computation of the transition manifold collective variables}
\label{algo:PointwiseRC}
\begin{algorithmic}[1]
\State Sample $M$ points $\{x_1,\ldots,x_L\}$ from $\X$ so that the metastable and transition regions are covered.
\State For each $x_i$, sample $p^\tau(x_i,\cdot)$ through parallel DNS.
\State Approximate the distance matrix $D\in \R^{L\times L}$,
$$
D_{ij} = d\left(p^\tau(x_i,\cdot), p^\tau(x_j,\cdot)\right).
$$
from the sampled $p^\tau(x,\cdot)$
\State Apply an unsupervised, distance-based manifold learning algorithm to $D$.
\Ensure Approximation to transition manifold collective variables $\xi^\ast$, evaluated at the points $\{x_1,\ldots,x_L\}$, i.e. 
$$
\{\xi^\ast(x_1),\ldots,\xi^\ast(x_L)\}~.
$$
\end{algorithmic}
\end{algorithm}

\subsection{State space sampling}
\label{subsec:state_space}

The evaluation points $\{x_1,\ldots,x_L\}$ should cover the dynamically relevant regions of $\X$, typically the metastable sets and the transition regions between them, evenly, in order for $\xi^\ast$ to capture the relevant dynamics globally. Various strategies exist, each with different advantages and disadvantages. While \emph{uniform sampling} of $\X$ may be the easiest to realize, it heavily oversamples the dynamically irrelevant regions, i.e., those with small $\pi$-measure. This holds true especially in high dimensions, where $\pi$ tends to be concentrated in a \mbox{(Lebesgue-)small} portion of $\X$. Hence, a large number of samples is required in order to cover the relevant regions with sufficient granularity. 

On the other hand, \emph{sampling from the invariant density $\pi$} solves this problem, but introduces a bias of oversampling the metastable regions (in which $\pi$ is heavily concentrated), while the transition regions are neglected. To alleviate this bias, the sampling of $\pi$ can undergo a second, subsampling step, through which the distribution is adjusted to yield more evenly-distributed samples. Two subsampling strategies were proposed in ref.~\cite{bittracherDatadrivenComputationMolecular2018}, namely the \emph{Poisson disk} and \emph{k-means subsampling}. Each of both strategies minimizes a specific approximation error of~$\xi^\ast$. Details on these algorithms can be also found in \cite{bittracherDatadrivenComputationMolecular2018}. In the later data analysis we will use k-means, mainly due to more readily available and robust numerical implementations.

\subsection{Transition density sampling}
\label{subsec:transition_density}

This step serves to create empirical approximations of the transition densities $p^\tau(x_i,\cdot),~i=1,\ldots,L$, as they are not available in closed form for all but the most simple dynamical systems. For this purpose, let $\Phi_\omega:\R^+\times \X\rightarrow\X$ denote the \emph{sample function} of the process $\{X_t\}$, where $\omega\in\Omega$ (the sample space underlying $\{X_t\}$). With Prob-distributed $\omega_1,\ldots,\omega_M$, this yields the \emph{empirical $M$-sample representative} of $p^\tau(x,\cdot)$:
$$
\{\Phi_{\omega_1}(\tau,x),\ldots,\Phi_{\omega_M}(\tau,x)\}\sim p^\tau(x,\cdot).
$$
If $\{X_t\}$ is characterized by an SDE with differentiable right hand side, $\Phi_\omega$ can be realized by a numerical integrator such as the Euler--Maruyama scheme, or a scheme of higher order. 

In the Voronoi-cell based discretization of a deterministic system $\Phi(\tau,\cdot)$, the randomness comes from the initial condition that is drawn from the Voronoi cell that the state~$x$ belongs to. The samples $\omega_i$ then refer to these randomly drawn initial conditions.

\begin{figure}
\centering
\includegraphics[width=0.4\textwidth]{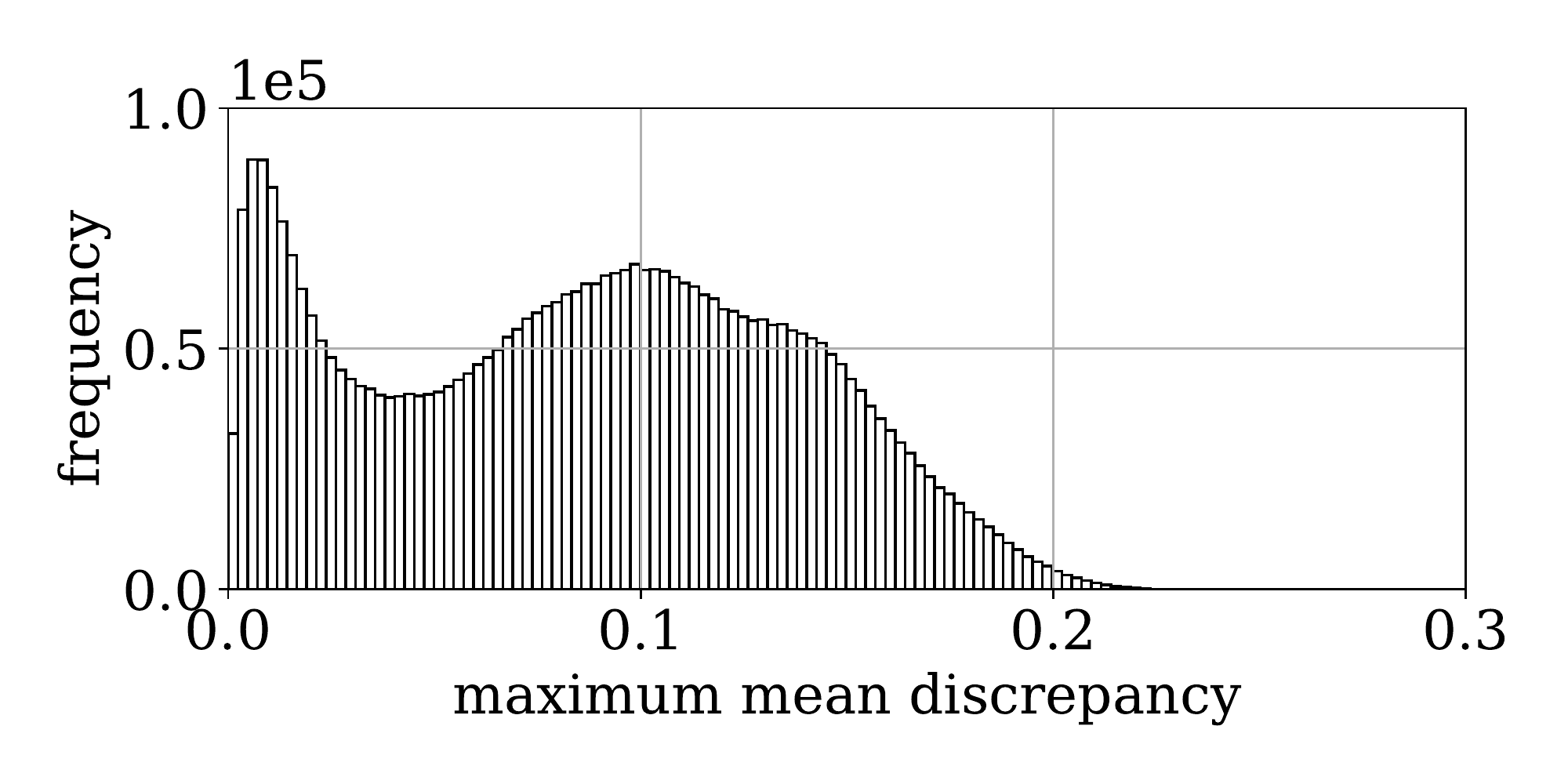}
\includegraphics[width=0.4\textwidth]{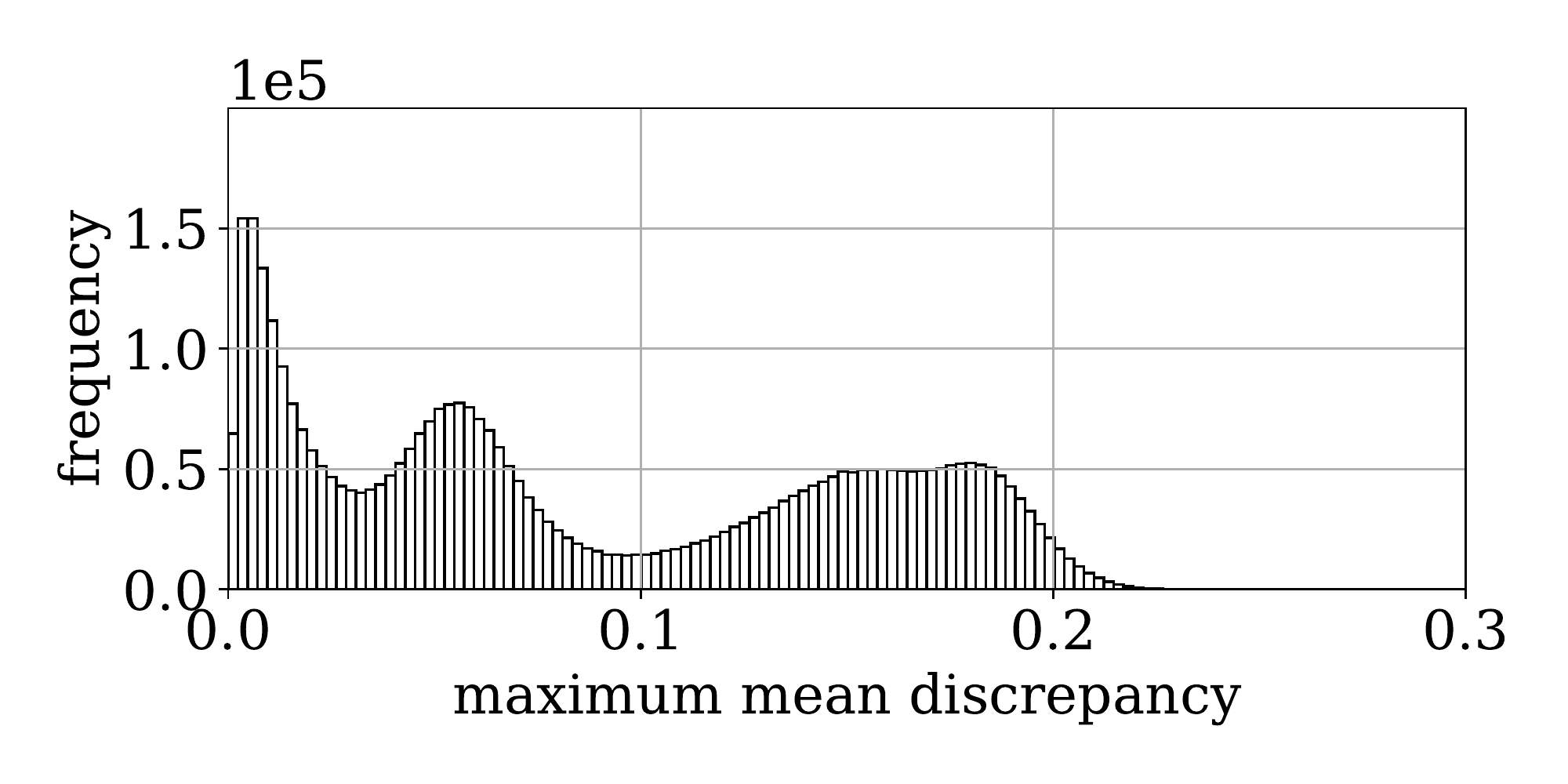}
\caption{Distribution of the maximum mean discrepancies between the embedded transition densities. We observe two accumulations of distances, one close to zero and one around 0.1, corresponding to intra- and inter-cluster distances, respectively. Data are for $Ra=10^6$ (top) and $10^7$ (bottom).}
\label{fig:kernelparameters}
\end{figure}

\subsection{Density distance matrix}
\label{subsec:MMD}

The assembly of the distance matrix $D$ serves only to prepare the data for the application of the manifold learning algorithm (see next subsection), but contains the crucial choice of the density metric $d:L^1(\X) \times L^1(\X) \rightarrow \R^+$, which determines both the  numerical feasibility of the algorithm and quality of its output. While 
$$
d( u, v) = \|u-v\|_{L^1}
$$
may seem like the obvious choice and the transition manifold framework is formulated with respect to this metric, it is notoriously hard to realize numerically (involving integrals over the $N$-dimensional space $\X$), and cannot be estimated directly from samples of $u,v$ (an approximate analytical expression of either $u$ or $v$ must be derived). Due to these difficulties, in \cite{BitEtAl17, bittracher_dimensionality_2020}, several alternative statistical distances were proposed, including the \emph{Maximum Mean Discrepancy} (MMD)~\cite{gretton_kernel_2012}, which will be used in the later data analysis. 

The MMD depends on the choice of a positive semi-definite kernel function $k:\X\times\X\rightarrow R^+$, and is analytically defined as
\begin{align*}
\label{eq:MMD}
d_{k}(u,v) &:= \| \mu(u) - \mu(v) \|_{\mathscr{H}(k)}, 
\intertext{where $\|\cdot\|_{\mathscr{H}(k)}$ is the norm induced by the inner product Here, $\mu$ denotes the \emph{kernel mean embedding}~\cite{smola2007hilbert} into~$\mathscr{H}(k)$. Its concrete form, as that of $\mathscr{H}(k)$, is irrelevant, since one has the following representation of inner products in $\mathscr{H}(k)$, readily giving a formula to compute the induced norm:}
\langle \mu(u), \mu(v) \rangle_{\mathscr{H}(k)} &= \int_\X \int_\X u(x) v(y) k(x,y)\ts d\pi(x)\ts d\pi(y).
\end{align*}
The Hilbert space $\mathscr{H}(k)$ induced by the inner product $\langle \cdot,\cdot\rangle_{\mathscr{H}(k)}$ is known as the \emph{reproducing kernel Hilbert space} (RKHS) associated with~$k$. There exists extensive literature on RKHS's and their functional analytical properties~\cite{hofmann2008kernel, berlinet2011reproducing}. Moreover, some of the authors have shown in~\cite{bittracher_dimensionality_2020} that the map $\M\mapsto \mu(\M)$ is an \emph{embedding}, i.e., preserves central topological properties of~$\M$. Importantly, the estimation of $\langle \cdot,\cdot \rangle_{\mathscr{H}(k)}$ and hence
$d_k( \cdot, \cdot)$ from samples is straight-forward: for
$$
\langle u,v \rangle_{\mathscr{H}(k)}^M := \frac{1}{M^2} \sum_{k,\ell=1}^M k(y_i^k, y_j^\ell),
$$
where $y_i^k\sim u$, $y_j^\ell\sim v$ holds 
$$
\langle \cdot,\cdot \rangle_{\mathscr{H}(k)}^M \rightarrow \langle \cdot,\cdot \rangle_{\mathscr{H}_k} \quad (M\rightarrow \infty).
$$

As described in the main text, we will compare the transition densities $p^{\tau}(x,\cdot)$ through their push forwards by the observation function $\psi = (\psi_1,\psi_2,\psi_3)^{\intercal}$, where $\psi_i$ denotes the $i$-th TICA coordinate. In other words, we need to compute the MMDs for the densities of $\psi(X)$ and $\psi(Y)$, where $X\sim p^{\tau}(x,\cdot)$ and $Y\sim p^{\tau}(y,\cdot)$, for different pairs~$x,y\in \X$. For this we chose the kernel $k(\xi,\eta) = \exp(-\|\xi-\eta\|^2 / \sigma)$ with~$\sigma=7$ and $\xi,\eta \in \mathbb{R}^3$ (since $\psi$ maps into~$\R^3$). The distribution of MMDs for both data sets is shown in Fig.~\ref{fig:kernelparameters}. The data for $Ra=10^6$ display a bimodal structure of the MMD histogram, which is going to show up in the discussion about bandwidth selection for the manifold learning method as well. The data for $Ra=10^7$ display a trimodal structure. Here we find that the first smallest scale is due to the strong clustering and disconnects the data set (cf.\ the discussion in section~\ref{sec:RBC_choices_bandidth}), even the second peak in the histogram indicates length scales that do not capture global properties of the data. It is the third local maximum of the MMD histogram that provides the right embedding.

\subsection{Manifold learning} 
\label{subsec: manifold learning}

We now apply a manifold learning method to the distance matrix $D$. Conceptually, this corresponds to learning an approximation to the combined parametrization and projection map $\mathcal{E}\circ\mathcal{Q}$ in Eq.~(6) in the main text. By pulling this map back to the starting points $x_i$, this yields the final collective variable $\xi^\star$. 

Our manifold learning method of choice is the \emph{diffusion maps algorithm}~\cite{coifmanDiffusionMaps2006}, as it has demonstrated good performance in practical scenarios in the past~\cite{bittracherDatadrivenComputationMolecular2018, bittracher_2021}, although other methods such as multidimensional scaling~\cite{young_multidimensional_2013}, local linear embeddings~\cite{roweis_nonlinear_2000}, or many others could be used as well.

We will not go into detail on the analytical derivation of diffusion maps here, but instead state only the algorithm. From the distance matrix $D\in\R^{L\times L}$, one first constructs a Markov transition matrix $M\in\R^{L\times L}$ via
\begin{equation}
M_{ij} = \frac{K_{ij}}{s_i},
\end{equation}
where $\smash{ K_{ij} = \exp\big(-D_{ij}^2 / \varepsilon\big) }$ is a \emph{similarity matrix} with some bandwidth parameter $\varepsilon>0$ (which should not be mixed with the kinetic energy dissipation rate field which is given in \eqref{ediss}) and~$\smash{ s_i= \sum_jK_{ij} }$. Being a Markov matrix, the leading eigenvalue of $M$ is 1 with corresponding constant eigenvector $\phi_1=(1,\ldots,1)$. The \emph{diffusion map}, hence our collective variable $\xi^\ast$ evaluated at the sample points $\{x_1,\ldots,x_L\}$, is now given by the following subdominant eigenvectors $\phi_2,\ldots\phi_{r+1}$. Here, the number $r$ can typically be determined by a gap in the spectrum of~$M$ or by plotting the dominant $\phi_i$ against one another and discarding the ones that do not carry additional geometric information, sometimes called higher-order harmonics.

We note that diffusion maps as developed in~\cite{coifmanDiffusionMaps2006} have a parameter $\alpha$ which is responsible for factoring out biases due to non-uniformity in the data sample density. The present simplified description corresponds to the case of $\alpha=0$. However, all our implementations use~$\alpha=1$, which is the value giving no dependence on sampling density in the limit $L\to\infty$. The coordinates~$\phi_i$ depend then on the geometric features of the underlying data manifold only.

\subsection{Bandwidth selection in diffusion maps}
\label{sec:bandwidth}

Let us briefly recapitulate a method for automatically determining a ``good'' kernel bandwidth parameter value~$\varepsilon$ in the diffusion maps method. The procedure stems from~\cite{CSSS08}, and has been later refined in~\cite{BeHa16}. A summary can also be found in~\cite[Appendix A.2]{Koltai_2020}.

Recall that the diffusion maps approach first turns pairwise distances in the data set $\{x_i\}_{i=1}^L$ into a similarity matrix $K \in \mathbb{R}^{L\times L}$ with entries $K_{ij}(\varepsilon) = \exp\left(-\varepsilon^{-1} D_{ij}^2\right)$. Note that we use the Gaussian kernel throughout this work. By averaging the entries of the similarity matrix, we define
\begin{equation}
S(\varepsilon) := \frac{1}{L^2} \sum_{i,j} K_{ij}(\varepsilon)\,.
\end{equation} 
We note the two limiting behaviors:
\begin{itemize}
\item
As $\varepsilon\to 0$, we have $K_{ij}(\varepsilon) \to \delta_{ij}$, the Kronecker delta. Thus, $S(\varepsilon) \to \frac1L$.
\item
As $\varepsilon\to \infty$, we have $K_{ij}(\varepsilon) \to 1$, thus $S(\varepsilon) \to 1$.
\end{itemize}
According to theoretical considerations, in between these two extremes there should be a region of affine-linear growth of $\log ( S(\varepsilon) )$ in~$\log (\varepsilon)$. This is suggested to be determined by maximizing
\begin{equation}
\label{eq:logderiv}
\frac{d \log(S(\varepsilon))}{d\log(\varepsilon)}\,,    
\end{equation}
with respect to~$\varepsilon$. The idea is that such an $\varepsilon$ is neither too small (compared with the data point density), nor too large in comparison to the diameter of the data point cloud. Such bandwidth parameters can then resolve the manifold structure of the data, if there is any, and represent it sufficiently well in the diffusion maps approach. In contrast, too small bandwidths tend to essentially disconnect the data set, while too large bandwidths disregard local geometric features by ``blurring'' them.

We note that the procedure also gives an estimate of the dimension of the data manifold. It follows as twice the value in~\eqref{eq:logderiv} for bandwidths $\varepsilon$ in the proper range. For strongly varying data density this procedure can give multiple ``optimal'' bandwidths $\varepsilon$ for different length scales, and one needs to compromise between resolution and connectivity (or otherwise break up the data in connected components and consider them one at a time). For even more irregular data one might need to use a \emph{variable bandwidth kernel}, as done in~\cite{BeHa16}.

\subsection{Bandwidth selection for the Rayleigh--B\'{e}nard convection data}
\label{sec:RBC_choices_bandidth}

\begin{figure*}[htb]
\centering
\includegraphics[width=0.32\textwidth]{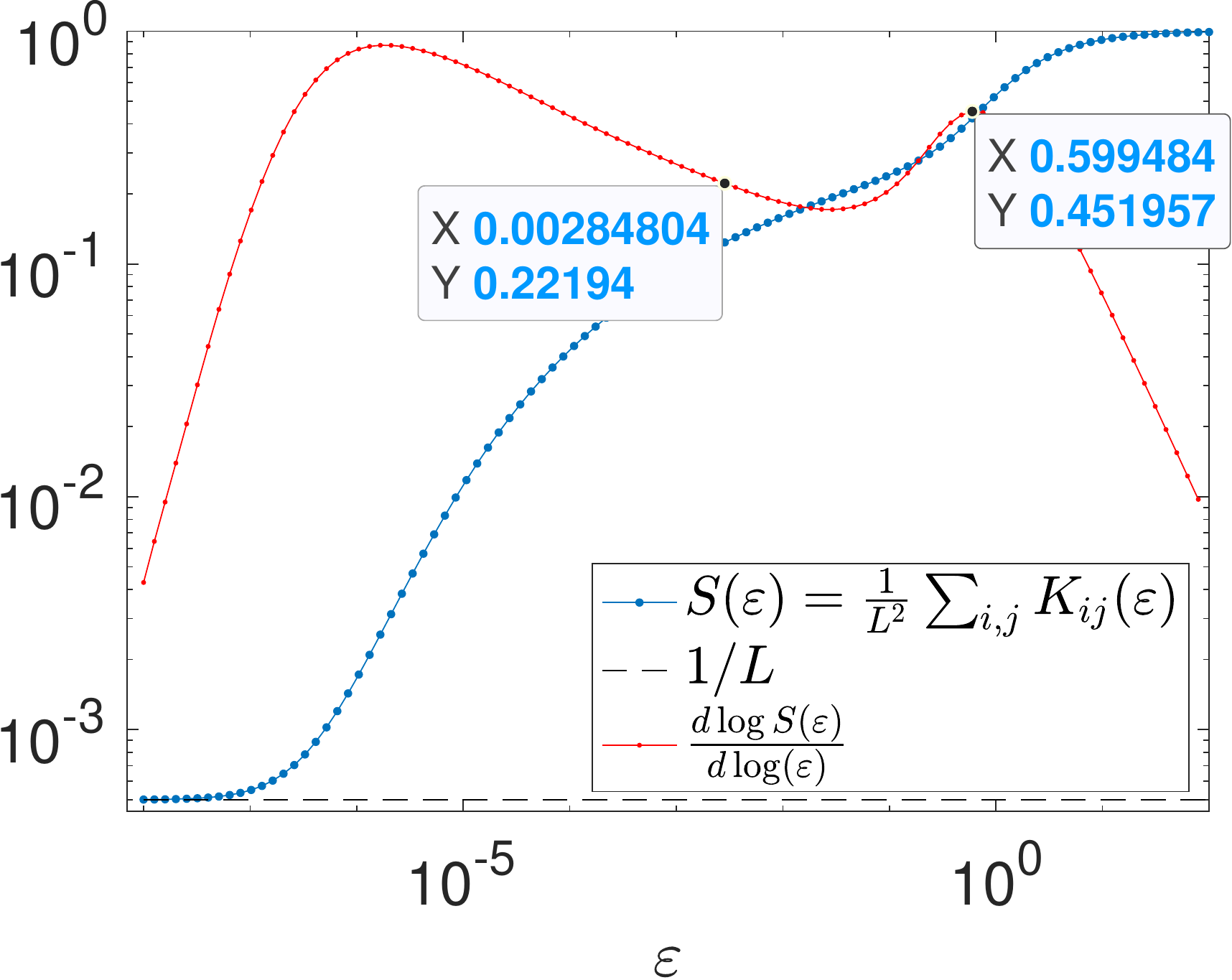}
\hfill
\includegraphics[width=0.32\textwidth]{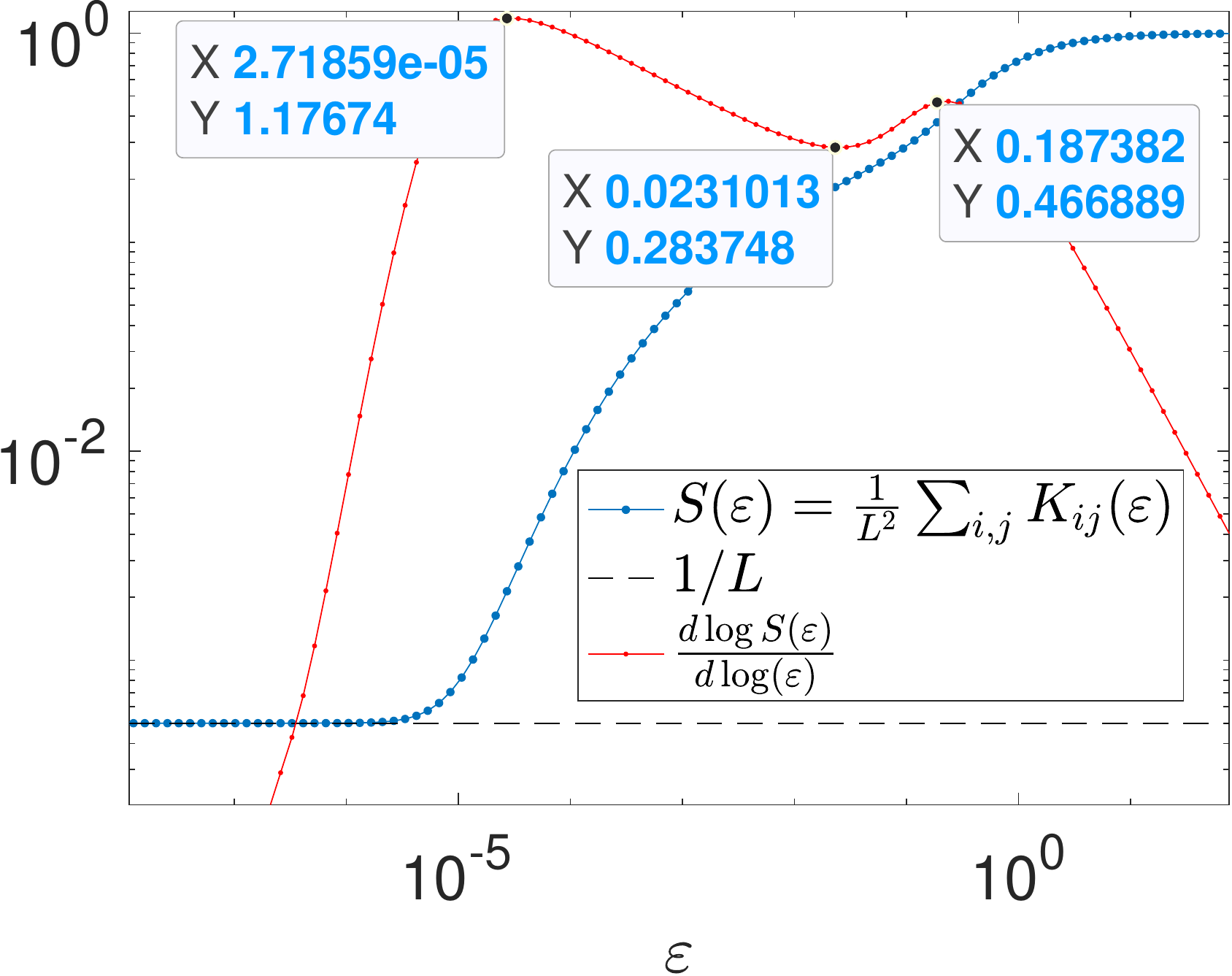}
\hfill
\includegraphics[width=0.32\textwidth]{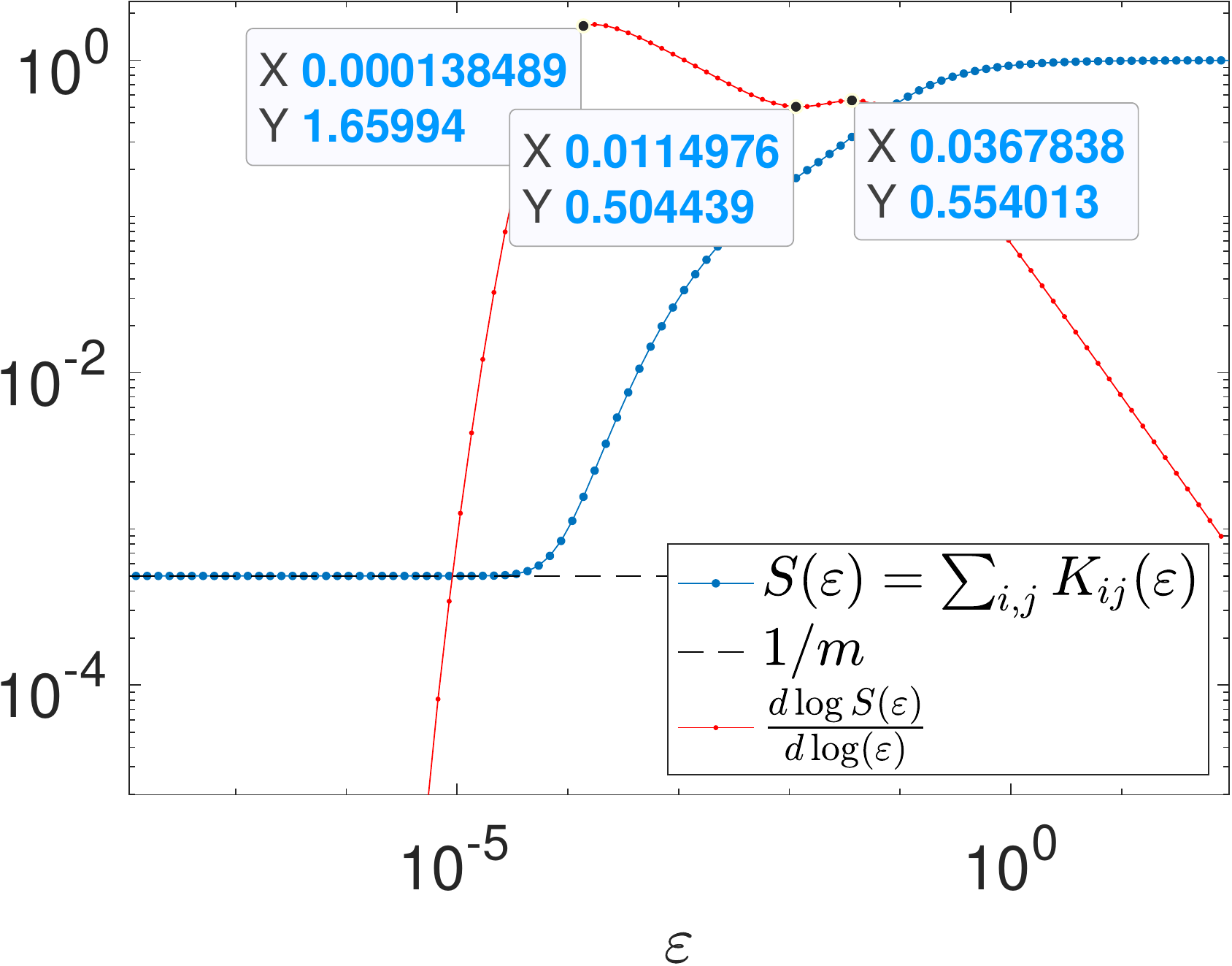}
\caption{The quantity $S(\varepsilon)$ and its double-logarithmic derivative for Maximum Mean Discrepancy distance matrices obtained from the first three (left), five (middle), and ten (right) TICA coordinates. The dashed line shows $1/L$, a lower bound for~$S(\varepsilon)$. The ticks in the plots show some values, for reference.}
\label{fig:RBC_epstuning}
\end{figure*}

To assess whether more TICA coordinates can capture more information about the geometric features of the transition dynamics of the RBC system, we ran the bandwidth-tuning analysis from appendix~\ref{sec:bandwidth} for distance matrices obtained from MMD computed from the first three, five, and ten TICA coordinates, respectively. The associated values of $S(\varepsilon)$ and the quantity~\eqref{eq:logderiv} are shown in Fig.~\ref{fig:RBC_epstuning} for $Ra=10^6$.

If the distance matrices would indicate an at most slowly varying data density, the logarithmic derivatives (red curves) would yield a single plateau, and every bandwidth $\varepsilon$ associated with the plateau would be an equally good choice for the representation of the data. The slight bimodal structure of these curves however indicates that the data set is somewhat irregular, and there are multiple proximity scales highlighting different features of the data set:
\begin{itemize}
    \item A smaller scale around the left-hand side maximum resolving local variability, but potentially disconnecting the data.
    \item A larger scale around the right-hand side maximum that views the data set as a connected manifold.
\end{itemize}

We conducted a detailed analysis for the data set obtained for the three dominant TICA coordinates. The reasons are that (i) essentially all three cases estimate the dimension of the data set to be at most two (cf. subsection~\ref{sec:bandwidth}), and that (ii) on suitable scales $\varepsilon$, the parametrizations of the three data sets are essentially equivalent (not shown). This indicates that the first three TICA coordinates already capture sufficient information on the transition statistics of the system, so that the addition of further TICA coordinates cannot add to this (at least not for the presently available RBC data). Moreover, auxiliary algorithms like $k$-means for the Voronoi center selection, are well-known to show deteriorating performance in high dimensions.

\begin{figure*}[htb]
\centering\includegraphics[width=0.99\textwidth]{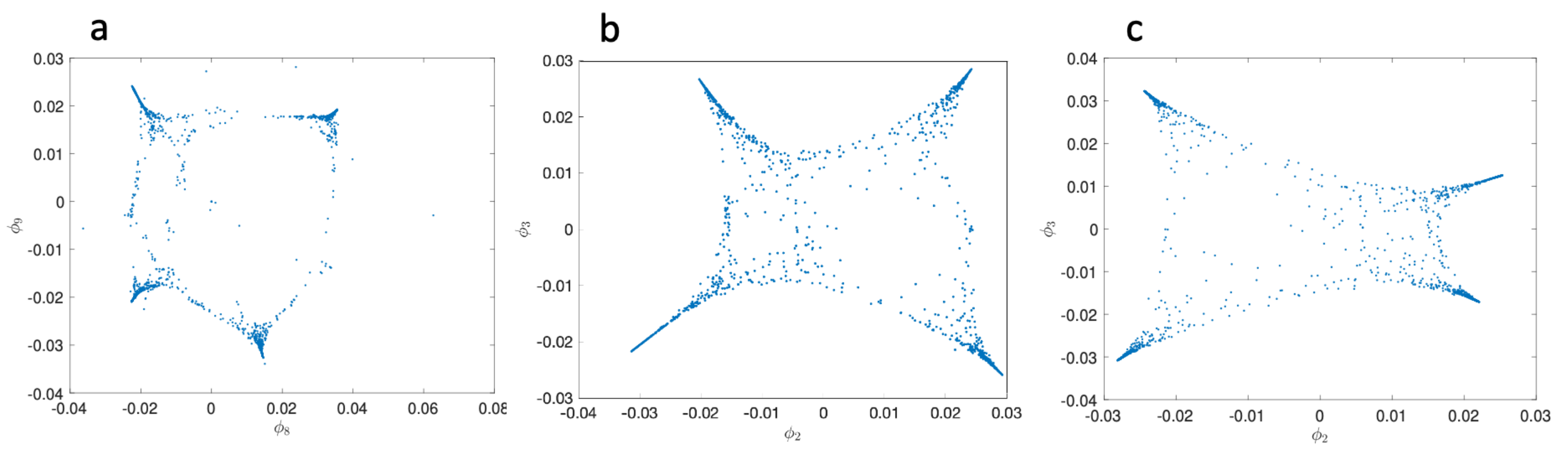}
\caption{Diffusion map paramterizations of the transition manifold for different bandwidths: $\varepsilon = 10^{-4}$ in (a), $3\times 10^{-3}$ in (b), and $2\times 10^{-2}$ in (c). The underlying Maximum Mean Discrepancy distance matrix was computed from the three dominant TICA coordinates. Note that since the diffusion map coordinates are computed as eigenvectors, their sign is not unique and may vary between different computations. Hence, to interpolate between the different panels, one might need to reflect one of the axes or both around zero.}
\label{fig:RBC_TICA03_dmaps}
\end{figure*}

Figure~\ref{fig:RBC_epstuning} left, corresponding to the distances from three TICA coordinates $(\psi_1, \psi_2,\psi_3)$, suggests to consider bandwidths $10^{-6} < \varepsilon < 0.5$. We observed that for $10^{-6} < \varepsilon < 10^{-4}$ the data set is essentially disconnected, i.e., the parametrization is dominated by single outliers and no structure involving multiple data points can be extracted. For $\varepsilon=10^{-4}$, the number of outliers reduces to a handful and the diffusion map eigenvectors $\phi_8,\phi_9$ already show some structure, see also Fig.~\ref{fig:RBC_TICA03_dmaps}(a).

Increasing the bandwidth, connectivity of the ``data manifold'' is achieved for $\varepsilon = 3\cdot 10^{-3}$, see Fig.~\ref{fig:RBC_TICA03_dmaps}(b), hence this is the bandwidth parameter which we use for further analysis in the main text. Figure~\ref{fig:RBC_TICA03_dmaps}(c) shows the diffusion map for $\varepsilon=0.02$, to show how a further increase in the bandwidth starts to blur local features and highlight global geometric aspects of the data set. This does not change if we would increase up to~$\varepsilon=0.5$ (not shown). Beyond this threshold, i.e., for very large bandwidths, the data set starts to look more and more point-like.

\bibliographystyle{unsrt}
\bibliography{MaityEtAl}

\end{document}